\documentclass[twocolumn]{aastex631}
\let\tablenum\relax
\usepackage{xcolor} % Usually not needed explicitly with AASTeX

\usepackage{graphicx}   % For including images
\usepackage{amsmath}    % For advanced mathematical typesetting
\usepackage{amssymb}    % For additional math symbols
\usepackage{siunitx}    % For SI units formatting
\usepackage{natbib}     % For bibliography management

\begin{document}

\title{Plasma instability in the front of ejected energetic electrons and Type III solar radiobursts}

%Plasma Instability at the Leading Edge of an Energetic Electron Beam and the Generation of Type III Solar Radiobursts

\correspondingauthor{Vladimir Krasnoselskikh}
\email{vkrasnos@gmail.com}

\author[0000-0002-6809-6219]{Vladimir Krasnoselskikh}
\affiliation{ LPC2E, OSUC, Univ Orleans, CNRS, CNES, F-45071 Orleans, France}
\affil{Space Sciences Laboratory, University of California, Berkeley, CA 94720, USA}

\author[0000-0002-0606-7172]{Immanuel Christopher Jebaraj}
\affil{Department of Physics and Astronomy, University of Turku, 20500 Turku, Finland}

\author{Tom Robert Franck Cooper}
\affiliation{LPC2E/CNRS, UMR 7328, 45071 Orléans, France}

\author[0000-0001-8307-781X]{Andrii Voschepynets}
\affil{Department of System Analysis and Optimization Theory, Uzhhorod National University, 88000 Uzhhorod, Ukraine}
\affil{Space Sciences Laboratory, University of California, Berkeley, CA 94720, USA}

\author[0000-0002-4401-0943]{Thierry Dudok de Wit}
\affiliation{International Space Science Institute, Bern, Switzerland}
\affiliation{ LPC2E, OSUC, Univ Orleans, CNRS, CNES, F-45071 Orleans, France}

\author[0000-0002-1573-7457]{Marc Pulupa}
\affil{Space Sciences Laboratory, University of California, Berkeley, CA 94720, USA}

\author[0000-0002-2011-8140]{Forrest Mozer}
\affil{Space Sciences Laboratory, University of California, Berkeley, CA 94720, USA}

\author[0000-0001-6427-1596]{Oleksiy Agapitov}
\affil{Space Sciences Laboratory, University of California, Berkeley, CA 94720, USA}
\affil{Astronomy and Space Physics Department, National Taras Shevchenko University of Kyiv, 03127 Kyiv, Ukraine}

\author[0000-0002-8110-5626]{Michael Balikhin}
\affil{University of Sheffield, Sheffield S10 2TN, United Kingdom}

\author[0000-0002-1989-3596]{Stuart D. Bale}
\affil{Physics Department, University of California, Berkeley, CA 94720-7300, USA}
\affil{Space Sciences Laboratory, University of California, Berkeley, CA 94720, USA}

%% Abstract
\begin{abstract} 

% \cvk{Type III radio bursts are a signature of the flux of near-relativistic electrons ejected during solar flares. These bursts are frequently observed by spacecraft, such as the Parker Solar Probe. It is traditionally believed that these electron beams generate Langmuir waves through two-stream instability, which are then converted into electromagnetic waves. In this study, we revise that model by examining how the electron distribution becomes truncated due to the ``time-of-flight" effect as the beam travels through an inhomogeneous plasma, such as the solar wind. Rather than the two-stream instability, this truncation destabilizes the distribution and leads to the generation of Langmuir waves. The instability grows until slower electrons arrive and dampen the waves. Our qualitative analysis shows that the resulting wave intensity growth and decay closely match the intensity-time profile of observed Type III radio bursts at the fundamental frequency, supporting this modified theory.}

Type III radio bursts are a signature of the flux of near-relativistic electrons ejected during solar flares. These bursts are frequently observed by spacecraft such as the Parker Solar Probe. It is traditionally believed that these electron beams generate Langmuir waves through the two-stream instability, which are then converted into electromagnetic waves. In this study, we revise that model by examining how the electron distribution becomes truncated due to the “time-of-flight” effect as the beam travels through a randomly inhomogeneous, and gently varying solar-wind plasma. Rather than the two-stream instability, this truncation destabilizes the distribution and leads to the generation of Langmuir waves via a linear instability; we deliberately confine our analysis to this linear regime and do not take into account the back reaction of the generated Langmuir waves on the electron distribution, which is nonlinear. The instability grows until slower electrons arrive and dampen the waves. Our qualitative analysis shows that the resulting wave intensity growth and decay closely match the intensity-time profile of observed Type III radio bursts at the fundamental frequency, supporting this modified theory.
\end{abstract}

%In radio spectrograms, these bursts appear as strong emissions that rapidly shift from high to low frequencies. The intensity of these bursts at each frequency rises quickly to a peak and then gradually decreases, showing an asymmetric profile.

%% Keywords

\section{Introduction} \label{sec:intro}

Type III radio bursts are the brightest radio emissions of solar origin and are signatures of fluxes of near-relativistic electrons ejected during solar flares and propagating away from the Sun. Their rate of occurrence and general characteristics during Parker Solar Probe's \citep[PSP;][]{Fox2016} close encounters with the Sun has been a subject of great interest \citep[e.g.,][]{Pulupa20,ChenL21,Jebaraj23c,Sishtla23,ChenL24,Pulupa24,Mozer24}. In the time-frequency spectrogram, they manifest as emission that drifts from high to low frequencies over a short period \citep{Suzuki85book}.

\begin{figure*}
    \centering
    \includegraphics[width=1\textwidth]{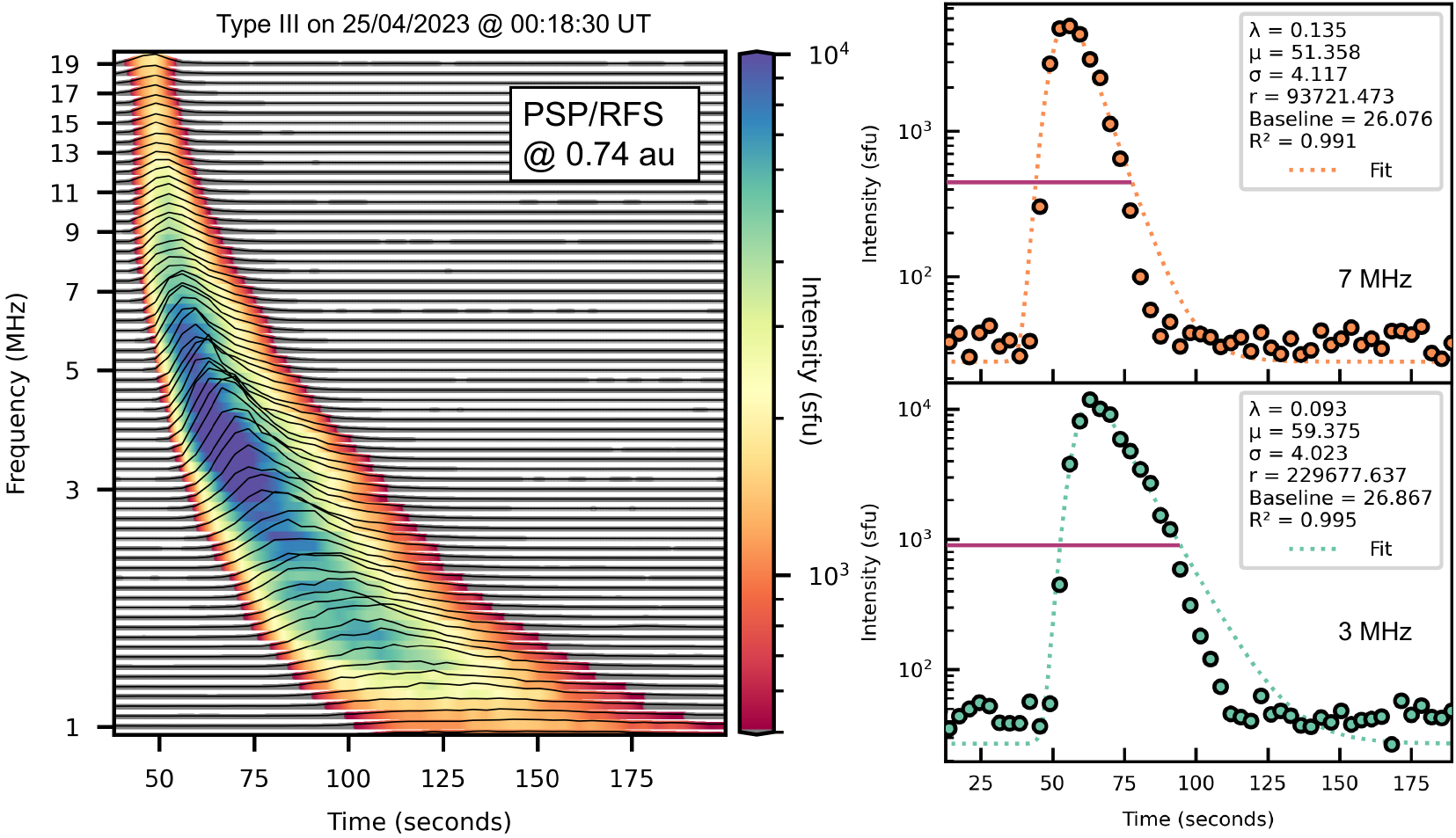}
    \caption{Time evolution of spectral features of a fundamental type III radio burst observed at various frequencies by PSP/RFS on April 25, 2023, at 00:18:30 UT is shown in the left panel. The right panels display time series at two frequencies: 7~MHz (top) and 3~MHz (bottom). Circle markers represent the data, which are fit with an exponentially modified Gaussian. The fit parameters, namely the mean (\(\mu\)), standard deviation (\(\sigma\)), exponential decay (\(\lambda\)), and the amplitude \(r\), are provided in the legend. The red horizontal lines in both the top and bottom panels mark the points beyond which the exponential function does not fit the decay of the signal.}
    \label{fig:spectrogram}
\end{figure*}

The process of generating Type III solar radio bursts is widely acknowledged to consist of two steps, as originally suggested by \cite{Ginzburg58} seven decades ago. According to their theory, the first step involves the excitation of Langmuir waves, which is generally attributed to the beam-plasma instability. The second step involves the transformation of wave energy from these electrostatic (ES) waves into electromagnetic (EM) emission generated at the fundamental frequency and its harmonic. The fundamental frequency or plasma frequency is given as \(\omega_\mathrm{p} = \sqrt{(n_\mathrm{e} e^2)/4\pi \epsilon_\mathrm{e} m_\mathrm{e}}\), where \(n_\mathrm{e}\) is electron density and \(\epsilon_\mathrm{e}\) is the permittivity of free space. It is well established that random density inhomogeneities are ubiquitous in the solar wind and are present across a wide range of spatial scales \citep{Celnikier83, Kellog05,ChenC13,Jebaraj_AGU2024}. Taking this into account, it was shown that the major process transforming ES energy into EM waves around \(\omega_\mathrm{p}\) is the direct partial transformation of Langmuir waves into ordinary EM waves during scattering on random density inhomogeneities \citep{Volokitin18, Krasnoselskikh19, Krafft22a}. As for the harmonic emission, it is thought to be generated as a result of the nonlinear coupling of primarily generated and backscattered/reflected Langmuir waves \citep{Tkachenko21, Melrose70, Willes1996}.

Recent observations of type III bursts by PSP, such as the one presented in Figure \ref{fig:spectrogram} suggest that they are bursty with rapidly varying features across different frequencies, indicating that these processes are transient and dynamic \citep{Jebaraj23, Jebaraj23c, Sishtla23, ChenL24}. An example of such a type III burst observed by the Radio Frequency Spectrometer \citep[RFS;][]{Pulupa17}, part of the FIELDS instrument suite \citep{BaleFIELDS} onboard PSP, is presented in Figure \ref{fig:spectrogram}. These bursts are often seen as fundamental-harmonic pairs \citep[with about 75\% probability;][]{Jebaraj23c}, but here we present a purely fundamental emission.  It is worth noting here that many of bursts registered by PSP are relatively weak, and are likely not observed at large distances of the order of 1 AU. The ridge-line plot also displays intensity-time profiles at frequencies from approximately 19~MHz to 1~MHz. At each frequency, the profiles are asymmetric, with EM waves quickly rising to a peak and then slowly decaying. Two single-frequency profiles are shown in the right panels, which are fit with an exponentially modified Gaussian with over 99\% accuracy. A detailed description of the fitting function can be found in Appendix \ref{App:EMG}. This indicates that the growth time (\(\tau_\mathrm{r}\)) follows a Gaussian function up to some peak, and then the signal decays (\(\tau_\mathrm{d}\)) at an exponential rate. Since EM waves are directly transformed from ES waves in the presence of density fluctuations, the EM wave profiles reflect the evolution of ES wave energy.

This leads to two important observational features that are not addressed by the widely accepted Ginzburg \& Zheleznyakov \citet{Ginzburg58} model. The first is the well-known asymmetry of type III time profiles \citep{Aubier72}. It is not clear to what degree the properties of the source electron distribution, and the growth of the waves contribute to this asymmetry. Recent observations from PSP \citep{Jebaraj23c} have shown that the more intense the emission is, the larger the asymmetry \((\tau_\mathrm{r}/\tau_\mathrm{d} \ll 1\)). This feature is intact even when the observer is at much greater distances from the Sun \citep[\(\sim 1.5\) AU,][]{Gerekos24} suggesting that the influence of the source is non-negligible. The second well-known feature that does not support the existing model is the rarity of observing the beam feature. Even though \cite{Lin81} identified the localized beam feature, only few follow-up experiments have detected beam features associated with Langmuir wave generation \citep{Lin81, Lin86}. Moreover, particle distribution measurements conducted during previous missions \citep[e.g. ISEE, STEREO][]{Ogilve77,Kaiser05} typically required much longer integration times than wave measurements. Consequently, these missions allowed only rough correlations between the arrival of energetic particles and the observed increases in remotely detected radio emissions or enhanced high-frequency wave activity. However, direct comparisons between simultaneous observations of energetic electron distribution dynamics and co-located wave dynamics were not feasible. This motivates a revision of the existing theory.

In order to address such behavior, the problem of ES wave generation must take into account the dynamics of localized processes at the front of the energetic electron flux. \cite{Lin81} first observed that wave activity develops around the front of the electron flux. These processes cannot be described by conventional beam-plasma interaction models or by spatial boundary problems for stationary processes. The problem must include transit-time effects of the beam, known as the “time-of-flight” phenomenon, which involves temporal variations of the particle source at each spatial location. This can be modeled with quasi-linear equations in a randomly inhomogeneous plasma, where the source term is crucial for system dynamics.

The idea that the propagation of the beam front may lead to wave activity in such a way that the beam particles generate waves, which are then absorbed by the slower part of the electron distribution, was first proposed by \citet{Zheleznyakov70} and \citet{Zaitsev74}. They considered how the local ES wave intensity rapidly increases and then slowly decreases as slower particles arrive and relax toward a plateau distribution. EM emission was then assumed to follow the evolution of ES Langmuir waves.

Here, we develop a very similar idea while adding an important element: the presence of random density fluctuations that crucially change the characteristics of the instability and the relaxation process. In order to do this, we aim to study the spatio-temporal evolution of a system with a localized source of energetic electrons and the instabilities that arise as they propagate away from the source. The time-of-flight mechanism of wave generation is based on a sharp increase in the electron distribution function due to the absence of beam particles with velocities below \(\frac{L}{t}\), where \(L\) is the distance from the source and \(t\) is the arrival time of energetic particles at the location where instability begins. We will also take into account the presence of random density fluctuations using the probabilistic model of beam-plasma interaction \citep[][]{Voshchep15a,Voshchep15b}, which accounts for the broadening of the Landau resonance due to phase velocity fluctuations of waves in a randomly inhomogeneous plasma while keeping the notion of the resonant velocity as the phase velocity without density fluctuations.

\begin{figure*}
    \centering
    \includegraphics[width=0.9\textwidth]{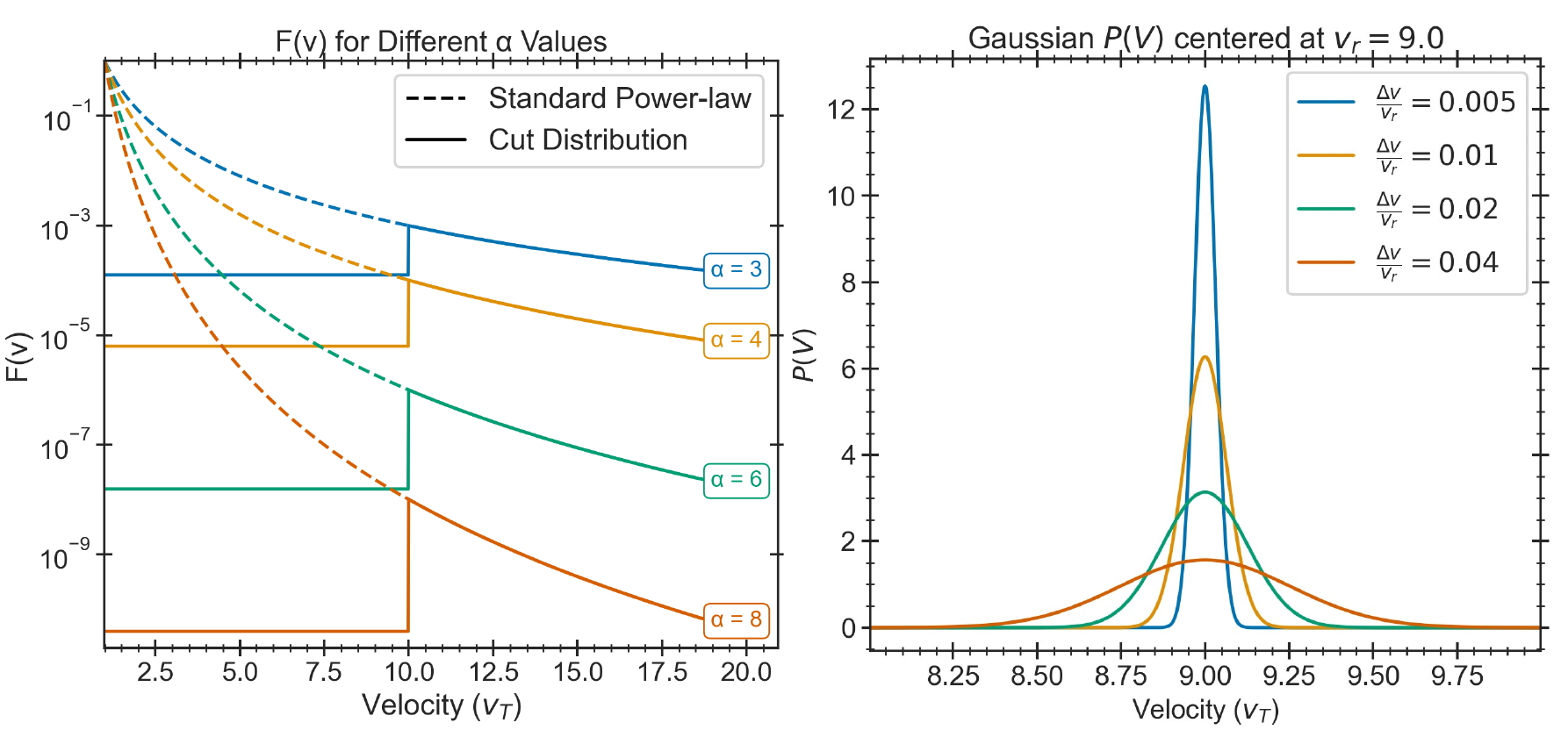}
    \caption{Illustration of distributions of energetic electrons and phase velocities under varying parameters. The left panel shows the truncated distribution of energetic electrons at position \(x = L\) for different values of the power-law spectral index (\(\alpha\)). The right panel displays the Gaussian distribution of phase velocities at a given \(L\) for various widths (\(\Delta V/V_\mathrm{r}\)) described by Equation \ref{eq_prob}. The \(P_{\omega}(V)\) is centered at \(V_\mathrm{r} = 9v_\mathrm{T}\). }
    \label{fig:Fig2}
\end{figure*}

\section{Formulation of the Problem}

For simplicity, let us consider the process of ejection of energetic electrons by a localized point source located at $x=0$. Assuming the presence of a weak magnetic field and neglecting the perpendicular motions of particles, one can formulate the problem as one-dimensional (1D). The background plasma density in our system monotonically decreases from some initial value $n_{e}(x=0)$ to zero, occupying the region from $x=0$ to infinity. It decreases according to some predefined law. Without loss of generality, one can suppose that the temperature of the plasma is constant. The density and temperature of the plasma determine the existence of plasma oscillations at any spatial location. The monotonic variation of the background plasma density is assumed to be “slow,” i.e., its characteristic length is much larger than the wavelength of the local waves; moreover, it is also much larger than the characteristic length of the instability development. This insight allows us to treat the waves locally. An important characteristic of our system is the presence of random density fluctuations everywhere. Our aim is to describe the interaction of local plasma waves with energetic electrons ejected from the source towards infinity. The injection process begins at the initial moment $t=0$. This system corresponds to the initial phase of ES wave generation by the flux of energetic electrons ejected during a solar flare. In the Ginzburg-Zheleznyakov type model, it corresponds to the first step in the two-step process of generating Type III radio bursts. Thus, the dynamics of the system will be described by the following set of equations, similarly to \citet{Voshchep15a} and \cite{Voshchep15b}:

\begin{equation}
\frac{\partial F}{\partial t} + v \frac{\partial F}{\partial x} =\sum_{i} \frac{2e^{2}}{m^{2}\omega} W_{i}(\omega) \frac{\partial}{\partial v} \left( P_{\omega}(v) \frac{\partial F}{\partial v} \right) + Q(v,r,t)  \label{eq1}
\end{equation}

\begin{equation}
\frac{dW_{i}}{dt} = \pi \omega_\mathrm{p} \frac{n_{b}}{n_\mathrm{e}}\int_{0}^{\infty} dV \, W_{i}(\omega) V^{2} P_{\omega}(V) \frac{\partial F}{\partial V}  \label{eq2}
\end{equation}

\begin{equation}
Q(v,r,t) = \Theta(t) \delta(x) n_\mathrm{b} \frac{(\alpha -1)}{v_{\min}} \left( \frac{v_{\min}}{v} \right)^{\alpha}, \quad \text{for } v > v_{\min}  \label{eq3}
\end{equation}
here $F$ is the energetic electrons distribution function, $n_\mathrm{b}$ is the beam density, ${v_{\min}}$ is the minimum velocity of energetic particle distribution, it is supposed however to be much larger than the thermal velocity of thermal electron distribution, $W_{i}$ is an energy density of the Langmuir wave with frequency $\omega$, $P_{\omega}$ probability distribution function for wave phase velocity at a given frequency. Equation \ref{eq1} describes interaction of the unstable electron population with a set of monochromatic waves. Hereafter we shall use $W$ instead of $W_{i}$ referring to a single Langmuir wave. 

Equations \ref{eq1} \& \ref{eq2} describe the generation of ES Langmuir waves and quasilinear relaxation of electron distribution function due to the action of these waves on the particles in the presence of the random density fluctuations. These fluctuations affect the phase velocity of waves that fluctuate as their phase velocity is dependent upon the local plasma density. More detailed description of the method of calculation of it and derivation of the functional dependence in the case of Gaussian distribution of the density fluctuations is presented in Appendix \ref{Appendix_B}.  
It is worth noting also that an important part of the description consists of the presence of the source of electrons described by the term $Q(v,r,t)$ on the right-hand side of Equation~\ref{eq1}. We suppose that at the initial time $t=0$, at point $x=0$, there is a source producing energetic electrons that begins to operate at $t=0$ and continues to do so thereafter. We choose the dependence of the distribution on velocity to be a power law with spectral index $\alpha$, which we consider as a parameter. These electrons begin to propagate towards $x > 0$. In simulations and observations, it is suggested that the source is rather a cloud of finite size, but if the time evolution of the generated wave packets is significantly smaller than the time of dispersion of the cloud, one can consider the source to function continuously after switching on. It is worth noting here that we suppose that \(v_\mathrm{T}\) is much smaller than even \(v_\mathrm{min}\) of energetic population, and that is why we neglect them in our further consideration.

To proceed to more detailed calculations, one should note that the distribution function of energetic electrons at point $x=L$ at a moment $t$ is found to be as follows:

\begin{equation}
F_\mathrm{b}(v,L,t) =
\begin{cases}
n_\mathrm{b}(v) \frac{(\alpha -1)}{v_{\min}} \left( \frac{v_{\min}}{v} \right)^{\alpha}, & \text{for } v \geq \frac{L}{t} \\
0, & \text{for } v < \frac{L}{t}
\end{cases}
\label{eq:truncated_distribution}
\end{equation}

As shown in the left panel of Figure \ref{fig:Fig2}, the distribution function of energetic electrons is truncated at a velocity separating particles that have arrived at the location from those that have not yet arrived. At point $x = L > 0$, plasma oscillations have a frequency close to $\omega_\mathrm{p}(L)$. In the left panel of Figure \ref{fig:Fig2}, we show the part of the distribution that is affected by this truncation for different distributions of the injected electrons given by the power‐law spectral index \(\alpha\). Considering that the distribution function at location $L$ remains the same as at the source, we restrict ourselves to distances that are not very large, since we do not take into account quasilinear evolution of the distribution function along its propagation. Rigorously speaking, our study is limited by the relaxation length of the beam–plasma interaction. However, such an approach may be validated by the majority of observations of energetic electron distributions, which are known to be of power‐law or double power‐law type. As our aim hereafter is to analyze the characteristics of the instability associated with time‐of‐flight–truncated electron distributions of the power‐law type, where the spectral index is one of the parameters in our study, we shall retain the power‐law index corresponding to the source distribution.  

For the sake of simplicity and transparency of the analysis, on the first step of our analysis we shall use the Gaussian probability distribution for the variations of the phase velocity associated with the density fluctuations as follows:

\begin{equation}
P_{\omega}(V) = \frac{1}{\sqrt{\pi} \Delta V} \exp \left[ -\frac{(V - V_\mathrm{r})^{2}}{(\Delta V)^{2}} \right],  \label{eq_prob}
\end{equation}
where $\Delta V$ is the width of the distribution, and $V_\mathrm{r}$ is the wave phase velocity in homogeneous plasma. The right panel of Figure \ref{fig:Fig2} shows the resulting Gaussian $P_{\omega}(V)$ centered at \(V_\mathrm{r} = 9v_\mathrm{T}\) for various $\Delta V$, normalized by $V_\mathrm{r}$. 

\begin{figure*}
    \centering
    \includegraphics[width=0.8\textwidth]{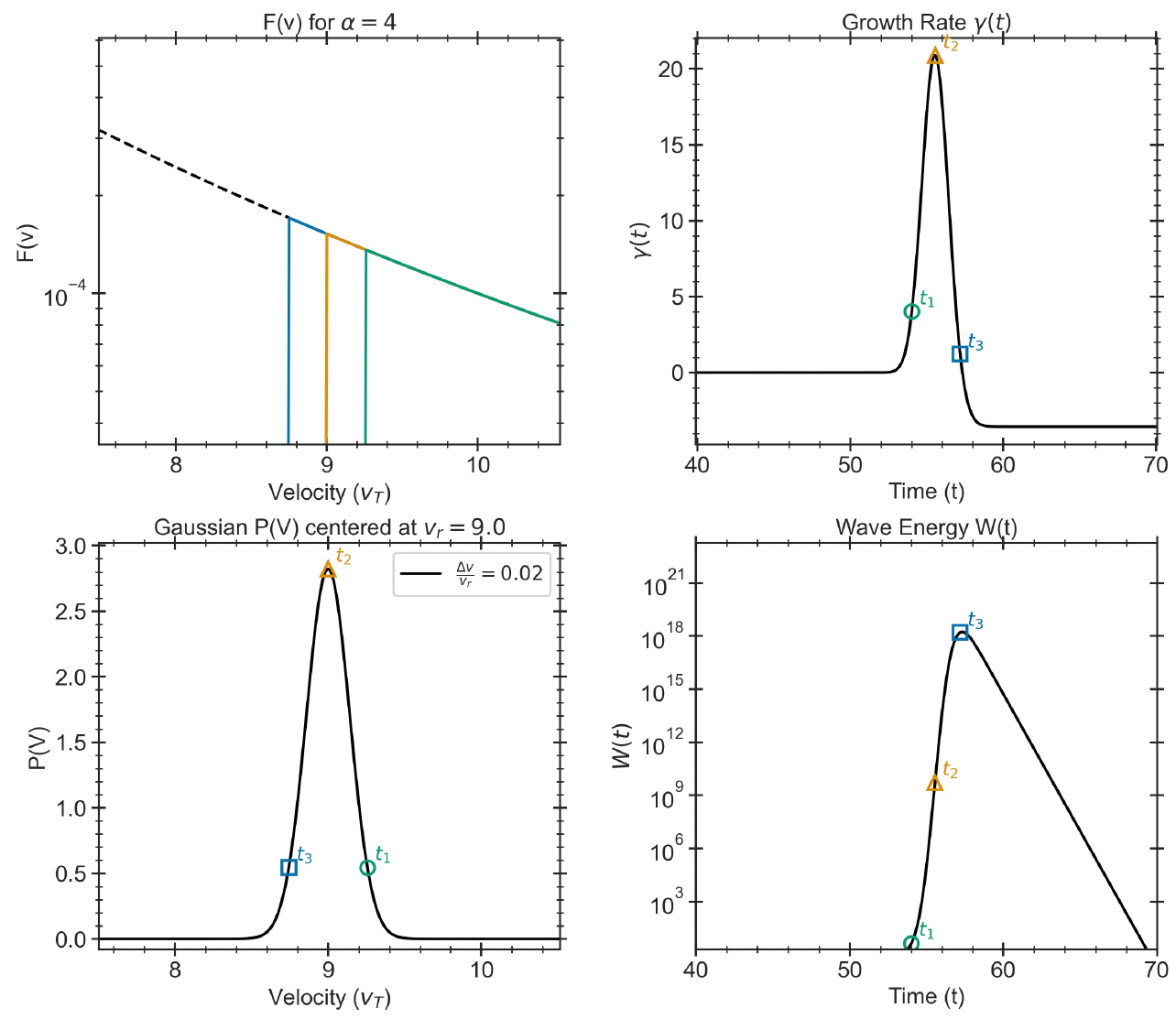}
    \caption{Instability of the truncated beam for Gaussian \(P_{\omega}(V)\) centered at $V_\mathrm{r} = 9 v_\mathrm{T}$, and $\alpha = 4$. Top-left panel shows the truncated distributions at three different times. Bottom-left panel shows the resulting probability distributions at these times, corresponding to the $10^{\text{th}}$ percentile ($t_3$, blue squares), $90^{\text{th}}$ percentile ($t_1$, green circles), and the velocity $V_\mathrm{r}$ itself ($t_2$, orange marker), which is the $50^{\text{th}}$ percentile. Top-right panel presents the growth rates $\gamma(t)$ obtained by solving Equation \ref{eq:gamma_analytic}, and the bottom-right panel shows the wave intensity $W(t)$ obtained using Equation \ref{eq:W}.}
    \label{fig:Fig3}
\end{figure*}

% \begin{figure}
%     \centering
%     \includegraphics[width=0.5\textwidth]{Figures/gaussian_pv.pdf}
%     \caption{Example Gaussian distribution function for different \(\Delta V\) based on Equation \ref{eq_prob}.}
%     \label{fig:Gaussian_distribution}
% \end{figure}

The increment of the instability for a wave with frequency $\omega$ in the randomly inhomogeneous plasma is given by:

\begin{equation}
\gamma(L,t) = \pi \omega_{\mathrm{p}0} \frac{n_\mathrm{b}}{n_\mathrm{e}} \int_{0}^{\infty} P_{\omega}(V) \frac{\partial F}{\partial V} V^{2} dV. \label{eq:gamma}
\end{equation}

Using the above-determined probability distribution and the distribution function for energetic electrons, one can find the following analytical expression for the increment:

\begin{align}
\gamma &= \pi \omega_\mathrm{p} \frac{n_\mathrm{b}}{n_\mathrm{e}} \left[ -\frac{\alpha (\alpha -1)}{v_{\min}^{2}} \int_{L/t}^{\infty} V^{2} \left( \frac{v_{\min}}{V} \right)^{\alpha +1} P_{\omega}(V) dV \right. \nonumber \\
&\left. + \left( \frac{L}{t} \right)^{2} P_{\omega}\left( \frac{L}{t} \right) \frac{(\alpha -1)}{v_{\min}} \left( \frac{v_{\min}}{L/t} \right)^{\alpha} \right]. \label{eq:gamma_analytic}
\end{align}

The resonant velocity for a Langmuir wave in the quasilinear theory of homogeneous plasma reads:

\begin{equation}
V_\mathrm{r} = v_\mathrm{T} \sqrt{ \frac{3}{2} \frac{\omega}{\omega - \omega_\mathrm{p}(L)} }. \label{eq:vr}
\end{equation}

\begin{figure*}[ht!]
    \centering
    \includegraphics[width=1\textwidth]{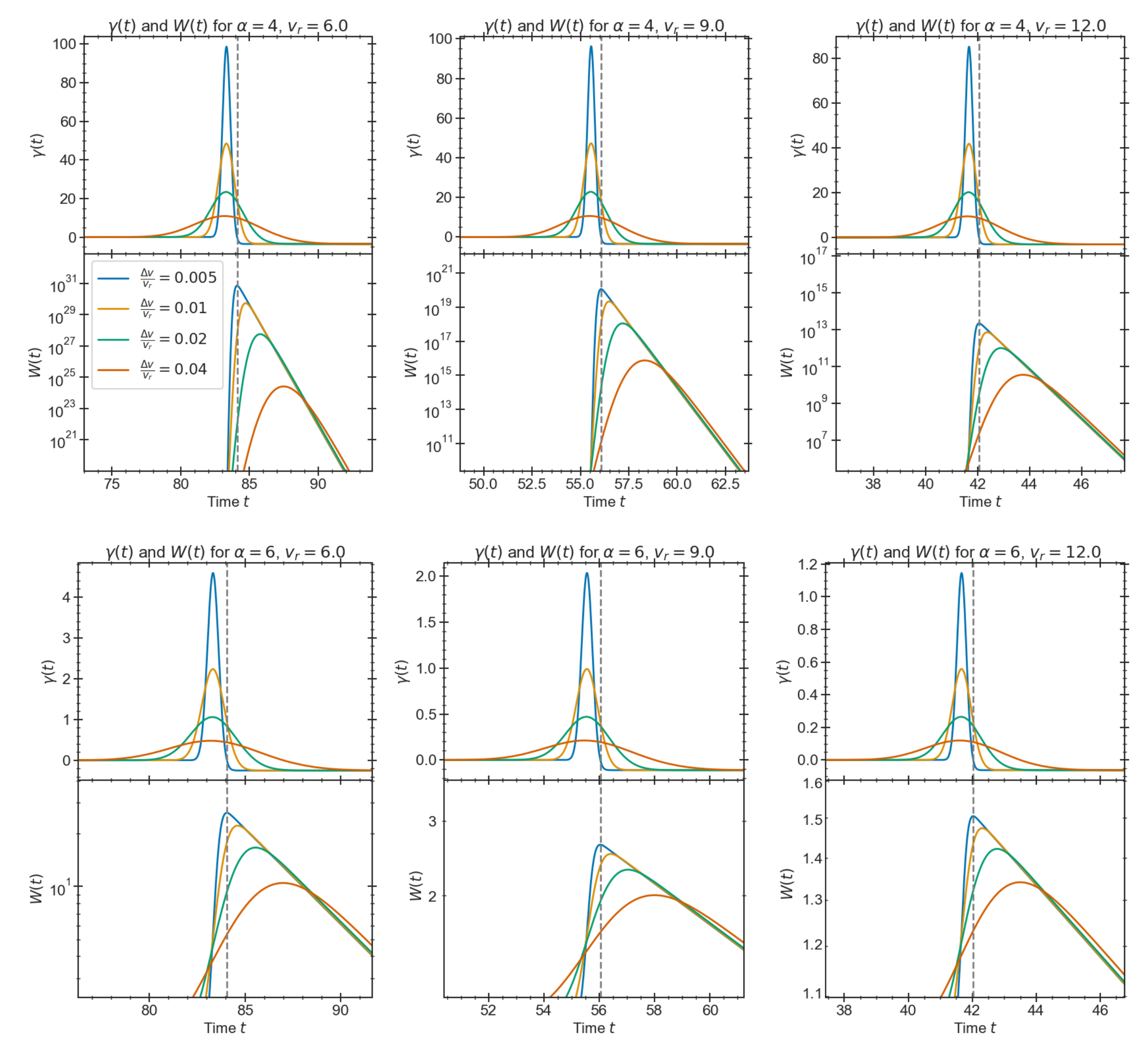}
    \caption{The growth (\(\gamma(t)\)) of the instability and the intensity \(W(t)\) of the generated waves for \(\alpha = 4\) (top row), and \(\alpha = 6\) (bottom row). Three different \(V_\mathrm{r}\) is used, namely, \(6 v_\mathrm{T}\) (left column), \(9 v_\mathrm{T}\) (middle column), and \(12 v_\mathrm{T}\) (right column). Each plot shows the effect of different \(\Delta V/V_\mathrm{r}\), where \(\Delta V\) is the width of the Gaussian \(P_{\omega}(V)\)). The vertical dashed line indicates the peak amplitude of the wave intensity corresponding to \(\Delta V/V_\mathrm{r} = 0.005\). }
    \label{fig:Fig4}
\end{figure*}

% \begin{figure*}[ht!]
%     \centering
%     \includegraphics[width=1\textwidth]{Figures/Gaussian_p2_final.pdf}
%     \caption{The growth (\(\gamma(t)\)) of the instability and the intensity \(W(t)\) of the generated waves for \(\alpha = 6\) (top row), and \(\alpha = 8\) (bottom row). Three different \(V_\mathrm{r}\) is used, namely, \(6 v_\mathrm{T}\) (left column), \(9 v_\mathrm{T}\) (middle column), and \(12 v_\mathrm{T}\) (right column). Each plot shows the effect of different \(\Delta V\) of the initial Gaussian \(P_{\omega}(V)\) chosen for the evaluation. The vertical dashed line indicates the peak amplitude of the wave intensity corresponding to \(\Delta V = 0.05\). }
%     \label{fig:wave_intensity_gaussian2}
% \end{figure*}

Although the source of high‐energy electrons may vary in space, the growth rate of the instability for a given plasma frequency and interaction frequency can be calculated locally at all times. As we discussed earlier, in our analysis nonlinear processes are neglected. In the simplified linear approximation, one can describe the evolution of the generated wave energy by the following expression:

\begin{equation}
W(L,t) = W_{0}(L) \exp {\left( \int_{0}^{t} \gamma(t')dt'\right)}, \label{eq:W}
\end{equation}
where $W_{0}(L,0)$ stands for the noise level of oscillations, which is supposed to be a stationary noise level.

The general formalism of the problem is illustrated clearly in Figure \ref{fig:Fig3}, which shows the instability arising from the truncation of the electron beam at a specific distance \(L\). For a given value of \(\alpha\), fewer electrons are found at higher velocities, which would result in the expected velocity dispersion due to their different flight times. By considering electron velocities corresponding to the \(10^{\text{th}}\) and \(90^{\text{th}}\) percentiles of the resulting Gaussian velocity distribution \(P_{\omega}(V)\), centered at a fixed velocity \(V_\mathrm{r}\), we observe that wave growth is highest at \(V_\mathrm{r}\) but is reduced by damping from slower electrons at the \(10^{\text{th}}\) percentile. Consequently, wave intensity peaks exactly at the point where damping begins to exceed growth. In Figure \ref{fig:Fig3}, this occurs at the arrival time of electrons at the \(10^{\text{th}}\) percentile.

\subsection{Qualitative Analysis of the Instability of the ``Time-of-Flight Electron Distribution''}

Let us begin by determining the conditions under which the instability can exist. First, we replace $\frac{L}{t}$ with the ``boundary velocity'' $U$. The expression for the increment becomes:

\begin{align}
\gamma &= \pi \omega_\mathrm{p} \frac{n_\mathrm{b}}{n_\mathrm{e}} \left[ -\frac{\alpha (\alpha -1)}{v_{\min}^{2}} \int_{U}^{\infty} V^{2} \left( \frac{v_{\min}}{V} \right)^{\alpha +1} P_{\omega}(V) dV \right. \nonumber \\
&\left. + U^{2} P_{\omega}(U) \frac{(\alpha -1)}{v_{\min}} \left( \frac{v_{\min}}{U} \right)^{\alpha} \right]. \label{eq:gamma_U}
\end{align}

The first term in this expression describes standard Landau damping as the function is monotonously decreasing. The second term represents a positive contribution defined by the step-like jump of the distribution from zero to a finite value. Taking into account the dependence of the probability distribution on $\Delta V$, one can notice that the positive part of the expression increases with the decrease of $\Delta V$. Hereafter, we shall limit our analysis to narrow distributions $P_{\omega}(U)$ where $\Delta V \ll V_\mathrm{r}$. The generation of waves due to the instability may exist under the condition that the second term is larger than the first one. The necessary condition for this requires that $\gamma$ should be positive at its maximum. The position of the maximum should satisfy the condition:

\[
\frac{d\gamma}{dU} = 0.
\]

There are two extrema of the function that correspond to:

\begin{equation}
U = \frac{V_\mathrm{r}}{2} \pm \sqrt{ \frac{V_\mathrm{r}^{2}}{4} + (\Delta V)^{2} }. 
\end{equation}

In our problem, only a positive value has physical meaning. One can show that this maximum increases with the decrease of $\Delta V$.  Taking into account the condition $\Delta V \ll V_\mathrm{r}$, one can find an analytical expression for the increment:

\begin{equation}
\gamma_{\max} = \pi \omega_\mathrm{p} \frac{n_\mathrm{b}}{n_\mathrm{e}} (\alpha -1) \left( \frac{v_{\min}}{V_\mathrm{r}} \right)^{\alpha -2} \left[ \frac{v_{\min}}{\sqrt{\pi} \Delta V} - \alpha \frac{v_{\min}}{V_\mathrm{r}} \right]. \label{eq:gamma_max}   
\end{equation}

It follows that there exists a threshold for the instability to occurexist:

\begin{equation}
    V_\mathrm{r} > \alpha \sqrt{\pi} \Delta V.  \label{threshold}
\end{equation}

This can be formulated either as the probability distribution for the phase velocity should be narrower than some limit or the spectral index of the power-law velocity distribution should be less than some critical value. The maximum is achieved at a speed value close to the resonant speed. Let us suppose that the condition $\Delta V \ll V_\mathrm{r}$ is satisfied and consider some dynamic features of the increment evolution under such conditions at a predetermined location $x=L$.

When $(U - V_\mathrm{r}) \gg \Delta V$, the increment is always smaller than the damping determined by the first integral term. However, the decrement remains very small as it is proportional to $P(U)$.

When $U$ approaches $V_\mathrm{r}$ at some critical velocity $(U_{\text{crit}} - V_\mathrm{r}) \sim \Delta V$, corresponding to the arrival at point $L$ of particles with velocities $U_{\text{crit}}$ and smaller, the increment begins to dominate. It grows with time and begins to increase rapidly when the velocity $U$ approaches $V_\mathrm{r}$. We are mostly interested in the time dependence of the increment at a predefined location $x=L$. As the maximum of the increment is achieved at some moment $t_{0}$, the major variation around the maximum will be described by the expression:

\begin{equation}
\gamma \simeq \gamma_{\max} \exp \left[ -\frac{(t_{0} - t)^{2}}{\tau_\mathrm{\gamma}^{2}} \right]. \label{eq:gamma_time}
\end{equation}

Here, $\tau_\mathrm{\gamma}$ may be interpreted as the rising time of the prompt increase of the increment, but one should note that the time of increase of the energy in Langmuir waves is determined by the characteristic rising time $\tau_{r}$
\begin{equation}
\tau_{r} \simeq \{(\gamma_\mathrm{max})^{-1}\}
\end{equation}
Since the transformation of Langmuir waves into EM waves around the fundamental frequency may be treated as having a very short time delay, it will correspond to the rising time used in data analysis of wave energy of EM waves at the fundamental frequency (see \citet{Jebaraj23}). Close to the peak, the rising time should be approximately the same as the time of decrease of the increment, but the wave energy still continues to grow. On this stage, the following description is applicable:

\begin{align}
W(L,t) &= W_{0}(L) \exp\left\{ \gamma_{\max} \int_{0}^{t} \exp\left[-\frac{(t_{0}-t')^{2}}{\tau_\gamma^{2}}\right] dt' \right\} \nonumber\\
       &\simeq W_{0}(L) \exp\left\{ -\gamma_{\max}\tau_\gamma \, \text{erf}\left(\frac{t-t_{0}}{\tau_\gamma}\right) \right\}. \label{eq:W_final}
\end{align}
where $\text{erf}\left(\frac{t - t_{0}}{\tau_\mathrm{\gamma}}\right)$ is the error function of the argument $\left(\frac{t - t_{0}}{\tau_\mathrm{\gamma}}\right)$. At a certain time (\(t_0\)), the wave intensity begins to decrease as the damping integral term overcomes the increment term.

\begin{figure*}
    \centering
    \includegraphics[width=1\textwidth]{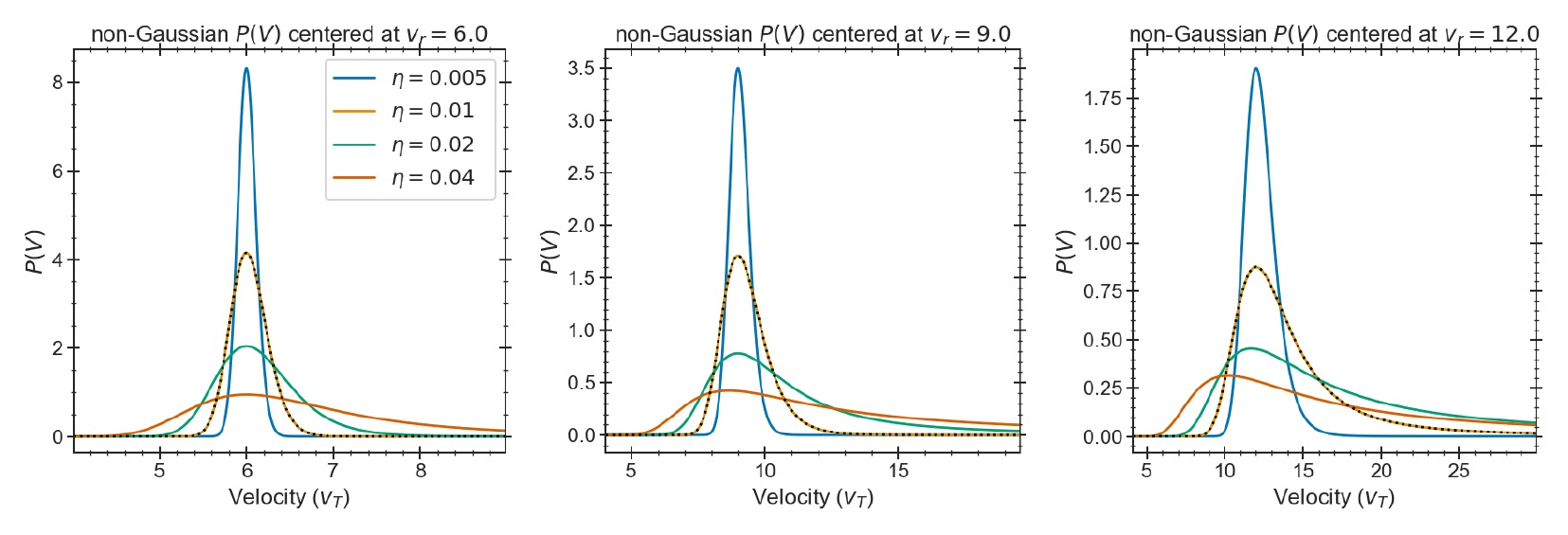}
    \caption{Non Gaussian \(P_{\omega}(V)\) for phase velocities given by Equation \ref{eq:non_GaussPv}. The three columns correspond to \(P_{\omega}(V)\) for three different \(V_\mathrm{r}\). Each panel shows the \(P_{\omega}(V)\) for four different values of \(\eta = \delta n/2n_\mathrm{e}\) which is the analog of \(\Delta V/V_\mathrm{r}\) in the Gaussian \(P_{\omega}(V)\). The dotted line correspond to the \(P_{\omega}(V)\) which considers only a single reflection given by Equation \ref{App:single_ref} only for \(\eta = 0.01\).}
    \label{fig:Fig5}
\end{figure*}

\begin{figure*}[ht!]
    \centering
    \includegraphics[width=1\textwidth]{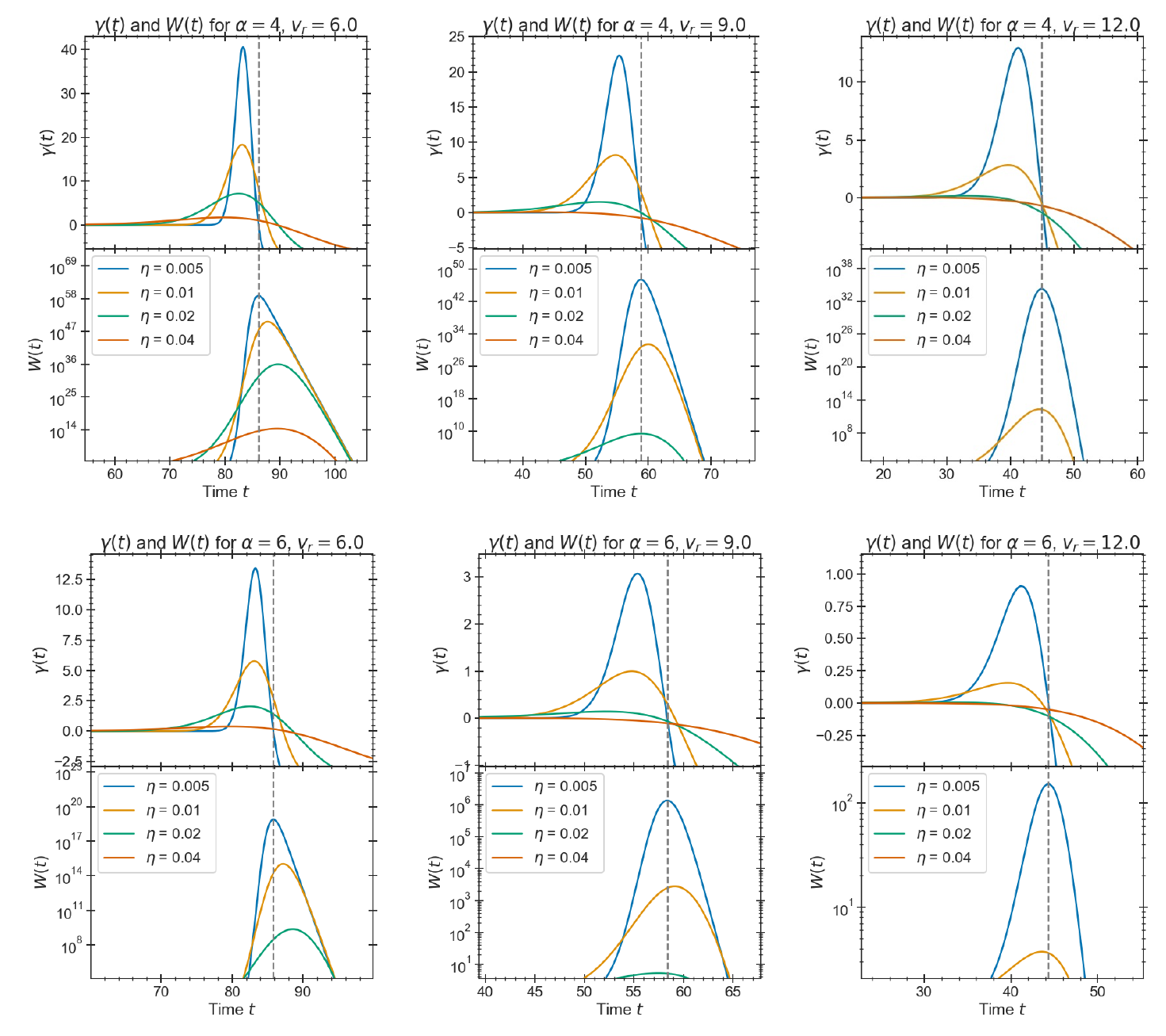}
    \caption{The growth (\(\gamma(t)\)) of the instability and the intensity \(W(t)\) of the generated waves for \(\alpha = 4\) (top row), and \(\alpha = 6\) (bottom row). Three different \(V_\mathrm{r}\) is used, namely, \(6 v_\mathrm{T}\) (left column), \(9 v_\mathrm{T}\) (middle column), and \(12 v_\mathrm{T}\) (right column). Each plot shows the effect of different \(\eta\) of the initial Gaussian \(P_{\omega}(V)\) chosen for the evaluation. The vertical dashed line indicates the peak amplitude of the wave intensity corresponding to \(\eta = 0.005\). }
    \label{fig:Fig6}
\end{figure*}

When the lower velocity particles arrive and the condition $(V_\mathrm{r} - U) > \Delta V$ begins to hold, the positive part of the increment becomes smaller than the damping rate determined by the first term, and the waves are absorbed by slower particles. The damping rate could be estimated as follows:

\begin{equation}
\nu = -\pi \omega_\mathrm{p} \frac{n_\mathrm{b}}{n_\mathrm{e}} \alpha (\alpha -1) \left( \frac{v_{\min}}{V_{\text{r}}} \right)^{\alpha -1}. \label{eq:nu} 
\end{equation}

It is interesting to note that the dynamics is described as a standard exponential decay with a constant damping rate. One can find that the ratio of the damping rate to the growth rate may be evaluated as:

\begin{equation}
\frac{\nu}{\gamma_{\max}} = \frac{\Delta V}{V_\mathrm{r}} \frac{\sqrt{\pi} \alpha}{\left[ 1 - \sqrt{\pi} \alpha \frac{\Delta V}{V_\mathrm{r}} \right]}. \label{eq:nu_gamma_ratio} 
\end{equation}

However, the application of this result is rather limited to just the initial phase of the signal decrease because, at the stage of signal decrease, the remotely observed EM waves do not represent a direct manifestation of the dynamics of Langmuir waves. The observations are ``polluted'' by additional effects due to wave escape from the generation region, propagation, and scattering of EM waves on inhomogeneities.

\section{Numerical Calculations of the Increment Evolution}

Hereafter, we supplement our qualitative analysis with more detailed numerical calculations of the increment to shed light on the dependence of the level of density fluctuations and the fine characteristics of energetic particle distributions. The level of density fluctuations is supposed to be equal to $\delta n$. The natural hypothesis is that the probability distribution of the density fluctuations is Gaussian. It is worth noting that if one has the functional dependence of the probability distribution of density fluctuations, the probability distribution of wave velocity may be found directly by means of the relation and their corresponding transformations.

Following the analysis described in detail in \citet{Voshchep15a} and presented hereafter in Appendix \ref{Appendix_B}, the probability distribution for the wave velocity may be found from the probability distribution for the density fluctuations:

\begin{equation}
    P_{\omega}(\delta n) \frac{d(\delta n)}{dV} = P_{\omega}(V),
\end{equation}
as per the conditions of resonance, taking into account that the wave frequency remains constant along the ray path. The velocity and density variation are related by the condition:

\begin{align}
\omega &\simeq \omega_\mathrm{p} \left(1 + \frac{3}{2} \frac{\omega^{2}}{V^{2}} \lambda_{D}^{2} + \frac{\delta n}{2n_\mathrm{e}} \right) \nonumber\\
&\simeq \omega_\mathrm{p} \left(1 + \frac{3}{2} \frac{v_\mathrm{T}^{2}}{V^{2}} + \frac{\delta n}{2n_\mathrm{e}} \right).
\end{align}

For a Gaussian distribution of density fluctuations, an expression for the probability distribution of velocities is rather cumbersome. Hereafter, one can  detailed derivation in Appendix \ref{Appendix_B}. Previous studies of the beam plasma interaction have shown that the instability development is most efficient for a narrow distribution of the probability distribution for velocity, which is quite close to a Gaussian distribution. In this section, we use the probability distribution for the wave phase velocity corresponding to the Gaussian distribution of density fluctuations. This will allow us to analyze a wider range of parameters and carry out an analysis for different values of the parameter $\alpha$ for the energetic electron distribution function.

Figure \ref{fig:Fig4} presents numerical results for two values of \(\alpha\) (\(4\) and \(6\) based on observations, e.g., \cite{Krucker07}) and three beam velocities \(V_{\mathrm{r}} = 6\,v_{\mathrm{T}}, 9\,v_{\mathrm{T}},\) and \(12\,v_{\mathrm{T}}\). We computed the wave growth rates \(\gamma(t)\) and the corresponding wave energy \(W(t)\) for four relative velocity spreads \(\Delta V / V_{\mathrm{r}} = 0.005,\;0.01,\;0.02,\) and \(0.04\). In all calculations, the beam truncation length was fixed at \(L = 500\), and the parameters \(\omega_{\mathrm{p}}\), \(n_{\mathrm{b}}\), and \(n_{\mathrm{e}}\) were each set to unity.

The characteristic time scale for beam–plasma interaction is related to the growth‐rate increment \(\gamma\). In homogeneous plasma, the relaxation time can be estimated as \citep{Voshchep15a}
\begin{equation}
    t \simeq \zeta \frac{\Lambda}{\gamma},
\end{equation}

where \(\zeta\) is of order \(3\)–\(10\) and \(\Lambda\) is the Coulomb logarithm. Normalizing time by \(\gamma^{-1}\) is therefore reasonable. Setting \(\pi\,(n_{\rm b}/n_{\rm e})\,\omega_{\rm p} = 1\) in dimensionless units is consistent with this choice.

The growth rates \(\gamma(t)\) are then normalized per particle by the factor \(V_{\mathrm{r}} / v_{\mathrm{ref}}\), with the reference velocity \(v_{\mathrm{ref}}\) fixed at \(6\,v_{\mathrm{T}}\). As expected, beams with higher \(V_{\mathrm{r}}\) initiate wave growth earlier, and the peak values of \(\gamma(t)\) decrease as \(V_{\mathrm{r}}\) increases. Steeper electron velocity distributions (\(\alpha = 6\)) exhibit significantly lower growth rates compared to shallower distributions (\(\alpha = 4\)).

Hereafter, we use the non-Gaussian probability distribution function of velocity corresponding to a Gaussian distribution of density fluctuations in the form:

\begin{align} 
P_{\omega}(V) &= \frac{1}{4} 
    \frac{\Big\{ \operatorname{erf} \Big( \frac{n_\mathrm{e}}{\delta n} \frac{3 v_\mathrm{T}^2}{V_\mathrm{r}^2} \Big) 
    - \operatorname{erf} \Big( \frac{3}{\eta} 
    \Big( \frac{1}{V_\mathrm{r}^2} - \frac{1}{V^2} \Big) \Big) \Big\}}{1 + \frac{1}{2} \Big[ 1 - \operatorname{erf} \Big( \frac{3}{\eta} \frac{1}{V_\mathrm{r}^2} \Big) \Big]} \nonumber \\
&\quad \times \Big\{ 1 + 
    \operatorname{erf} \Big[ \frac{3}{\eta} \Big( \frac{1}{V_\mathrm{r}^2} - \frac{1}{V^2} \Big) \Big] \Big\}.
\end{align} \label{eq:non_GaussPv}

where $\eta = \frac{\delta n}{2n_\mathrm{e}}$, corresponding to the density fluctuation level. This \(P_{\omega}(V)\) is close to a Gaussian distribution on velocities when

\[
\frac{\delta n}{n_\mathrm{e}} \ll 3 \frac{v_\mathrm{T}^{2}}{V_\mathrm{r}^{2}}.
\]

This last limit was used in our qualitative description of the instability evolution in the previous section. It is worth noting that when $\eta \rightarrow 0$, the distribution becomes a Dirac delta function $\delta(v - V_\mathrm{r})$, as it should in a homogeneous plasma. Figure \ref{fig:Fig5} shows the velocity distribution \(P_{\omega}(V)\) for three values of \(V_\mathrm{r}\) and four values of \(\eta\). As previously mentioned, \(P_{\omega}(V)\) closely resembles a Gaussian distribution when \(\eta\) is small. However, three key deviations from Gaussian behavior become apparent as \(\eta\) and \(V_\mathrm{r}\) increase: (1) \(P_{\omega}(V)\) becomes increasingly asymmetric; (2) the distribution width expands significantly beyond a Gaussian profile; and (3) the peak of \(P_{\omega}(V)\), expected at the chosen \(V_\mathrm{r}\), shifts toward lower velocities (e.g., at \(V_\mathrm{r} = 6\) in the first panel).

These characteristics are clearly reflected in Figure \ref{fig:Fig6}, which presents \(\gamma(t)\) and \(W(t)\) for similar numerical parameters as Figure \ref{fig:Fig4}. The peak shift in \(P_{\omega}(V)\) results in the corresponding \(\gamma(t)\) shifting to later times, while the broader \(P_{\omega}(V)\) distribution facilitates wave growth across a wider velocity range. Although overall trends remain similar to the Gaussian scenario, wave amplitudes for \(\alpha = 4\) become notably large at lower velocities (\(V_\mathrm{r} = 6\)). This significant amplitude increase suggests the rising importance of nonlinear processes under these conditions. Detailed examination of these nonlinear effects, however, is beyond this paper's scope and will be pursued in future research.

Finally, since the non-Gaussian \(P_{\omega}(V)\) calculations include multiple reflections (see Appendix \ref{Appendix_B}), we briefly investigate the impact of single versus multiple reflections from density humps. The dotted line in Figure \ref{fig:Fig5} shows \(P_{\omega}(V)\) for a single reflection case at \(\eta = 0.01\). The difference between single and multiple reflections is minor for the parameters considered here, as their ratio is roughly on the order of \(10^{-2}\). Nevertheless, the difference becomes more noticeable at higher values of \(V_\mathrm{r}\) and for larger \(\eta\) (not shown here).

\begin{figure*}[ht!]
    \centering
    \includegraphics[width=1\textwidth]{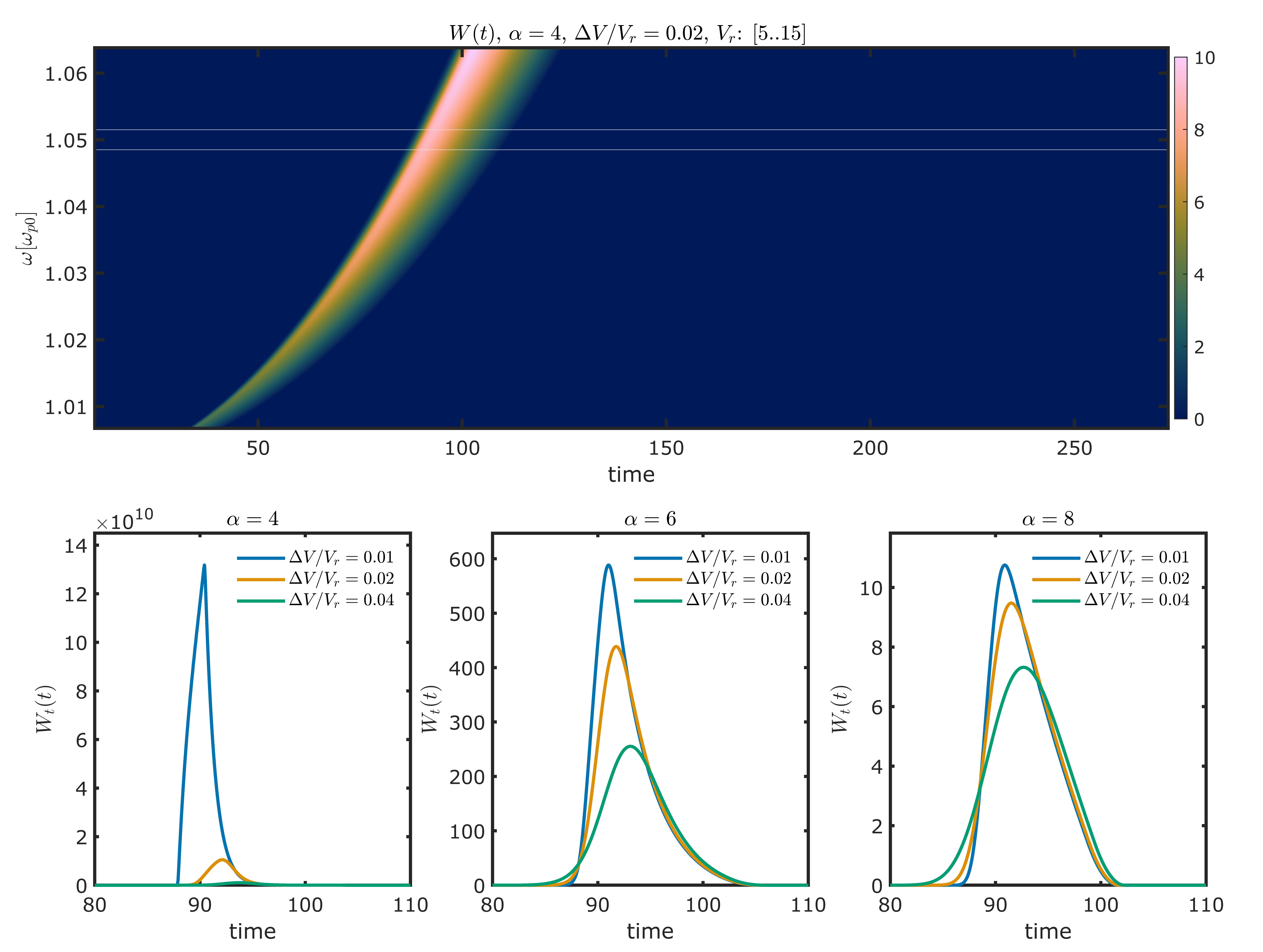}
    \caption{Evolution of the spectrum of the Langmuir waves at $L=500$ obtained using equations \ref{eq:gamma_analytic} and \ref{eq:W} for $\alpha=4$ and $\Delta V/ V_{\mathrm{r}}=0.02$ (top panel). The spectrum consists of the waves with $V_{\mathrm{r}}$ that lie in the range $5$ to $15 v_\mathrm{T}$. Bottom panels: temporal evolution of the total wave energy density close to $\simeq 1.05 \omega_\mathrm{p}(L)$, obtained by integration of the spectrum within the frequency range denoted by two horizontal lines in the top panel. Left to right: $\alpha=4$, $\alpha=6$, $\alpha=8$. Different colors correspond to different $\Delta V/ V_{\mathrm{r}}$. }
    \label{fig:Fig7}
\end{figure*}

\section{Discussion \& Conclusions}

The primary characteristics of type III solar radio bursts include rapid variations in their emission frequency peak over time, which is determined by the propagation of energetic electrons ejected from the Sun outward into the solar wind in the anti-sunward direction. The decrease in solar wind density with distance causes the characteristic emission frequency to decrease accordingly. The original mechanism for generating type III bursts was proposed by \cite{Ginzburg58} and involves two main steps. Initially, ES Langmuir waves are generated through beam-plasma instability. Subsequently, these ES waves undergo nonlinear transformation into EM waves at fundamental and harmonic frequencies.

Studies of density fluctuations in the solar wind have clarified their critical role in generating primary waves and their subsequent transformation into EM emission. Consequently, describing the instability requires incorporating the effects of random density fluctuations, which we address using a probabilistic model of beam‐plasma interactions. While it is widely accepted that primary Langmuir waves arise from the beam‐plasma instability (also known as the bump‐on‐tail instability), we propose an alternative mechanism. In our model, however, the primary instability arises from a truncated electron distribution formed at the leading edge of the injected energetic electron flux due to the time‐of‐flight effect. An important feature of the distribution functions of energetic electrons measured in the solar wind is the absence of any positive slope \citep{Krucker07,Lorfing23}, which excludes the possibility of generating waves via the conventional beam‐plasma instability \footnote{The work of \cite{Lorfing23} analyzes spectra obtained at peak flux. To study the positive slope, the instantaneous spectra must be examined.}. An alternative mechanism may be provided by truncated electron distributions caused by the sharp front of the propagating flux of energetic electrons. In this paper, we study the possible characteristics of such an instability. 

The question arises as to how these distributions form and what determines their characteristics. It is widely accepted that electron distributions in the solar wind are best described by kappa functions. Kinetic models of the solar wind trace back to the pioneering works of Scudder \citep{Scudder92, Scudder13} and Lemaire and co‐authors \citep{Lemaire96, Lemaire97, Lemaire2010}, which incorporate the filtering of the electron distribution by the electrostatic potential established around the chromosphere and low corona. Such filtering naturally truncates the electron distribution, since the positively charged Sun prevents part of it from escaping into the heliosphere. Energetic electrons are ejected into interplanetary space during strong perturbations in active regions, when reconfigurations of magnetic and electric fields produce energetic electron fluxes. It is reasonable to assume that this ejection significantly alters the kappa‐type electron distribution by depleting its high‐energy tail. Because the tails of kappa distributions are often well approximated by power laws, analyzing the instabilities associated with these escaping tails is certainly justified.

We analyzed the instability using a power-law electron distribution characterized by several different spectral indices $\alpha$, and various resonant wave velocities \(V_\mathrm{r}\). The main characteristics of this instability are as follows:
\begin{itemize}
    \item The instability strengthens for lower (or shallower) \(\alpha\) due to a slower decrease in particle number at higher velocities.
    \item The instability is stronger for lower velocities because the jump in the distribution function is larger.
    \item The instability becomes stronger as the probability distribution of wave velocity narrows. We demonstrated this behavior explicitly for a Gaussian probability distribution of wave phase velocity. Furthermore, when employing a more complex distribution associated with Gaussian density fluctuations, we confirmed that the instability is stronger at lower fluctuation levels, corresponding to narrower velocity probability distributions. Furthermore, we note a shift towards lower \(V_\mathrm{r}\) with increasing fluctuations.
    \item The instability is slightly stronger when wave reflections are few (single-reflection approximation), though the difference from multiple reflections is relatively minor for small \(V_\mathrm{r}\).
\end{itemize}

These characteristics of the instability of truncated electron distributions due to the time‐of‐flight effect naturally explain the temporal variations in wave amplitude observed at a given frequency. This applies even in the linear approximation, which may be valid only for weak bursts. However, the majority of the bursts registered onboard PSP are weak, and many of them may not be visible at a distance of 1 AU. At any given location, there may exist a spectrum of waves corresponding to different resonant velocities. The spectrum's evolution initially exhibits rapid growth, peaks, and then declines due to damping by slower electrons arriving later. In Figure \ref{fig:Fig7}, we illustrate the temporal evolution of waves corresponding to resonant velocities ranging from $15\, v_\mathrm{T}$ to $5\, v_\mathrm{T}$. Our analysis remains within a linear approximation; therefore, certain nonlinear saturation effects are beyond our current description and will require further study.Nonetheless, the dynamic features we predict for weaker bursts agree well with the morphology of type III bursts in observational data, its asymmetry. This is particularly so for steeper values of \(\alpha\), the difference between which is shown in the bottom panels of Figure \ref{fig:Fig7}. These two issues were discussed recently by \citep{Jebaraj23c} who found that the asymmetry was only dependent on the intensity of the emission and there were no particular dependence upon frequency. It is clear from our results that the asymmetry is a localized feature of the emission process and is only dependent upon the width of the distribution (or level of density fluctuations). However, our model has limitations in describing the decay phase of bursts beyond some level, as the EM wave spectrum at the fundamental emission frequency is affected by scattering and other propagation effects \citep{Steinberg71}.

In this study, we did not consider nonlinear effects - particularly the back-reaction of waves on particles. Numerical simulations have shown that these effects are crucial for the relaxation of the electron distribution function in randomly inhomogeneous plasmas \citep{Krafft13,Krafft15,Krafft16,Voshchep15a,Voshchep15b}.

The model introduced in this paper highlights an essential distinction between the linear instabilities underlying type III and type II radio bursts. While electron distributions in both types result from a time‐of‐flight effect, this term refers to different processes in each case. In type II bursts, it denotes the spatially stationary reflected electron population around a curved shock front, whereas in type III bursts it describes the nonstationary electron injection at the leading edge of an energetic electron flux.

Although not extensively explored, the gradient of the thermal velocity \(v_\mathrm{T}\) with respect to distance from the solar corona provides important insights into two key aspects: (1) the initiation height of type III solar radio bursts, and (2) the impact of wave interactions on the initial power‐law electron distribution. The initiation height is particularly significant because it serves as a remote sensing diagnostic for energetic particle releases during solar flares and other eruptive events \citep{Kouloumvakos15, Mitchell25}. Specifically, the gradient in \(v_\mathrm{T}\) implies that, in the lower corona—where \(v_\mathrm{T}\) is relatively high—the instability is primarily driven by electrons with higher energies. Furthermore, previous studies have shown that beam–plasma relaxation processes in randomly inhomogeneous plasmas modify not only the slower portion of the electron distribution, as predicted by classical quasi‐linear theory, but also transfer significant wave energy to higher‐energy electrons \citep{Krafft13, Voshchep15a}. This effect is relevant to the instability discussed here because it substantially alters the initially injected electron distribution. Both effects challenge the assumption of a beam‐type quasi‐monoenergetic electron population responsible for type III bursts, indicating that beam characteristics near the Sun differ considerably from those observed in situ. A combination of these factors may explain observed delays between energetic particle fluxes and type III burst emissions \citep{Mitchell25}, and could also introduce spectral breaks in near‐relativistic electron distributions detected by spacecraft \citep{Krucker07, Strauss20, Dresing21, Jebaraj23b}. These predictions remain qualitative but establish a foundation for future quantitative analyses.

In summary, random density fluctuations in the solar wind significantly modify the conventional two-step generation theory initially proposed by Ginzburg \& Zheleznyakov. The primary revision involves changing the underlying linear instability: it is neither bump-on-tail nor traditional beam-plasma instability, but rather an instability of truncated energetic electron distributions caused by the time-of-flight effect. An additional significant modification involves directly converting energy from primarily generated Langmuir waves into EM waves at plasma frequencies due to wave reflection off random density fluctuations \citep{Hinkel92,Krasnoselskikh19}. As for the generation of the second harmonic emission, it remains unchanged from the original Ginzburg \& Zheleznyakov model -- it occurs via coupling between primary and reflected Langmuir (or Z-mode) waves \citep{Melrose17, Willes1996, Tkachenko21}.

%Such modifications could manifest as spectral breaks in the energetic electron distributions observed by spacecraft \citep{Krucker07,Dresing20,Jebaraj23b}.
\section*{acknowledgements }

This research was supported by the International Space Science Institute (ISSI) in Bern, through ISSI International Team project No.557, “Beam-Plasma Interaction in the Solar Wind and the Generation of Type III Radio Bursts”. 
V.K. and I.C.J., were supported through the Visiting Scientist program of the International Space Science Institute (ISSI) in Bern.
Parker Solar Probe was designed, built, and is now operated by the Johns Hopkins Applied Physics Laboratory as part of NASA’s Living with a Star (LWS) program (contract NNN06AA01C). Support from the LWS management and technical team has played a critical role in the success of the Parker Solar Probe mission.
V.K., acknowledges financial support from CNES through grants, ``Parker Solar Probe", and ``Solar Orbiter", and by NASA grant 80NSSC20K0697. I.C.J., is grateful for support by the Research Council of Finland (SHOCKSEE, grant No.~346902), and the European Union’s (E.U's) Horizon 2020 research and innovation program under grant agreement No.\ 101134999 (SOLER). The study reflects only the authors' view and the European Commission is not responsible for any use that may be made of the information it contains.

\appendix

\section{Exponentially modified Gaussian} \label{App:EMG}

In this study, similar to \cite{Jebaraj23c} and \cite{Gerekos24}, we use an exponentially modified Gaussian (EMG) function to fit the time-frequency signals of type III radio emissions. The EMG is obtained by convolving a Gaussian with mean \(\mu\) and standard deviation \(\sigma\) with an exponential decay of rate \(\lambda\). For a time variable \(t\), the EMG is defined as

\begin{equation}
    f(t) \;=\;
\int_{-\infty}^{t}
\frac{1}{\sqrt{2\pi}\,\sigma}\,
\exp\!\left[-\frac{1}{2}\left(\frac{\tau - \mu}{\sigma}\right)^2\right]
\,\lambda\,e^{-\lambda\,(t-\tau)}\,d\tau,
\end{equation}

which can be expressed in closed form using the error function,

\begin{equation}
    f(t) \;=\;
\text{baseline}
\;+\;
r\,\frac{\lambda}{2}\,
\exp\!\Biggl[\frac{\lambda}{2}\Bigl(2\mu + \lambda\sigma^2 - 2t\Bigr)\Biggr]
\Biggl[
1 - \mathrm{erf}\!\left(\frac{\mu + \lambda\sigma^2 - t}{\sqrt{2}\,\sigma}\right)
\Biggr].
\end{equation}

Here, \(\mu\) and \(\sigma\) determine the position and spread of the Gaussian peak, \(\lambda\) controls the rate of the exponential tail, \(r\) scales the amplitude, and \(\text{baseline}\) provides a vertical offset.

\vspace{0.5em}

\noindent To understand the EMG quantitatively and why it is preferred for fitting the type III time-frequency signals (Figure \ref{fig:spectrogram} right panels), let us consider its two limiting components. First, a simple Gaussian,

\begin{equation}
     g(t) = \exp\!\left[-\frac{(t - \mu)^2}{2\,\sigma^2}\right],
\end{equation}
   
is perfectly symmetric and decays as \(\exp\!\bigl(-\tfrac{(t-\mu)^2}{2\sigma^2}\bigr)\). Its decay is quadratic in the exponent, meaning that any asymmetry (or tail) in the data is not captured.
    
\noindent Secondly, a pure exponential,
    \begin{equation}
        e(t) = e^{-\lambda t},
    \end{equation}
    
decays as \(\exp(-\lambda t)\), which is linear in the exponent. This produces an inherently skewed function (with a skewness of 2 for a standard exponential) but lacks a natural peak.

The EMG effectively combines these two behaviors by having a Gaussian-like rise and an exponential fall. Analytically, the difference is clear, that is, while the Gaussian decays as \(\exp\!\left[-\tfrac{(t-\mu)^2}{2\sigma^2}\right]\), the exponential decay is \(\exp(-\lambda t)\). By mixing these two forms, the EMG accommodates a finite, symmetric rise (characterized by \(\mu\) and \(\sigma\)) and a gradual, asymmetric decay (governed by \(\lambda\)).

This dual nature can be quantified in practical peak fitting. A Gaussian fit tends to underestimate the trailing edge, leading to large systematic residuals in the tail region. An exponential model, when forced to represent a peaked structure, overestimates the signal near the maximum and fails to capture the proper curvature. In contrast, the EMG fit often yields lower least-squares residuals, reduced \(\chi^2\), and higher \(R^2\) values, thereby minimizing the sum-of-squares error.

Furthermore, the parameters of the EMG provide direct quantitative insight. In particular, the product \(\lambda\sigma\) serves as an indicator of skewness—a larger \(\lambda\sigma\) means a more pronounced exponential tail relative to the Gaussian core. This clear, analytically tractable parameterization makes it straightforward to distinguish between the quadratic decay of a Gaussian and the linear decay of an exponential.

\section{Probability distribution for wave velocity in randomly inhomogeneous plasma} \label{Appendix_B}

The lowest order effects describing wave--particle interaction take into account the most important process, namely, resonant interaction of particles and waves having the same velocities (i.e. when the phase velocity of the wave is exactly equal to the particle velocity). Quasilinear theory in plasma and trapping of particles by a finite amplitude wave are two limits of this approach. The basic theory is well developed and applied to different plasma systems in homogeneous plasmas. However, when the plasma is randomly inhomogeneous and the phase velocity depends on plasma parameter fluctuations (as is the case for Langmuir waves in plasma with density fluctuations), the resonant interaction becomes strongly perturbed since the interaction must now account for fluctuations of the wave phase velocity. In addition, the wave may interact not only with the particles that would be in resonance in a homogeneous plasma but also with other particles as the local velocity varies; thus, on some intervals the wave is found to be in resonance with particles that are locally resonant.

In order to describe this phenomenon we introduce the notion of a probability distribution of the wave velocity for a wave having frequency \(\omega\) in a plasma where the average plasma frequency (when density fluctuations are absent) is \(\omega_{\mathrm{p}0}\). The most transparent way to derive the probability distribution is to discretize the spatial interval into finite but small segments and to calculate the probability that the wave velocity equals the particle velocity inside a given interval. The size of the interval should satisfy the following conditions: it should be much larger than the wavelength of the wave (so that the notion of phase velocity is meaningful) and it should be significantly smaller than the distance over which the particle distribution function changes appreciably, i.e.,

\[
\lambda \ll L \ll 10\,\frac{V_{r}}{\gamma}\,.
\]

We assume that the profiles of velocity and density inside the interval are both monotonic and may be approximated by straight lines between the endpoints. To evaluate the probability that the phase velocity is equal to the particle velocity on the interval

\[
l_i < x < l_{i+1}\,,
\]

(where \(l_i\) and \(l_{i+1}\) are the beginning and end points of the interval) one should suppose that at one end the velocity is larger than the wave velocity and at the other it is smaller, i.e.,

\[
\min(u_i, u_{i+1}) < V < \max(u_i, u_{i+1})\,.
\]

Since the interval is small and the difference between \(u_i\) and \(u_{i+1}\) is small, the probability can be written as

\[
P(u_k<V<u_j)=P(u>V)\cdot P(u<V)\,.
\]

This form holds if one does not take into account the effect of wave reflection. Let the probability of a single wave reflection be denoted by \(P_\mathrm{ref}\). In the limit where the reflection probability is small, the normalization is written as

\begin{equation} \label{App:single_ref}
    1 = P(u>V)+P(u<V)+P_\mathrm{ref}\,.
\end{equation}

It corresponds to the limit when one considers that the wave reflection occurs only once. In this limit the probability for velocity distribution is presented by above written expression. 
If the reflection probability becomes relatively large, one must consider multiple reflections. In the limit of an infinite number of reflections the total probability that the wave keeps its initial direction is estimated as

\begin{align}\label{Appendix_A_E6}
P_{\mathrm{initial}} &= (1-P_\mathrm{ref}) + P_\mathrm{ref}^2(1-P_\mathrm{ref}) + P_\mathrm{ref}^4(1-P_\mathrm{ref}) + \cdots \nonumber\\[1mm]
  &= (1-P_\mathrm{ref})\sum_{n=0}^{\infty}\left(P_\mathrm{ref}^2\right)^n \nonumber\\[1mm]
  &= \frac{1-P_\mathrm{ref}}{1-P_\mathrm{ref}^2} = \frac{1}{1+P_\mathrm{ref}}\,.
\end{align}

and correspondingly, the probability that the wave reverses its direction is

\begin{equation}\label{Appendix_A_E7}
P_{\mathrm{ref}} = \frac{P_\mathrm{ref}}{1+P_\mathrm{ref}}\,.
\end{equation}

Let us now evaluate these probabilities for a Gaussian distribution of density fluctuations given by

\begin{equation}\label{Appendix_A_E8}
P(\delta n) = \frac{1}{\sqrt{\pi}\,\delta n}\exp\!\left[-\frac{\delta n^2}{n_{e}^2}\right]\,.
\end{equation}

The frequency of the propagating wave does not vary along the wave path. In an inhomogeneous plasma it is determined by

\begin{equation}\label{Appendix_A_E9}
\omega = \omega_\mathrm{p}\left(1+\frac{3}{2}k(x)^2\lambda_D^2+\frac{1}{2}\frac{\delta n(x)}{n_\mathrm{e}}\right)
=\text{const}\,.
\end{equation}

where \(k(x)\) is the local wave vector and \(\delta n(x)\) is the local density fluctuation. Under the resonant condition for the wave phase velocity this relation can be re-written as

\begin{equation}\label{Appendix_A_E10}
\omega = \omega_\mathrm{p}\left(1+\frac{3}{2}\frac{v_\mathrm{T}^2}{V^2}+\frac{1}{2}\frac{\delta n(x)}{n_\mathrm{e}}\right)\,.
\end{equation}

To determine the integration limits it is useful to note that

\begin{equation}\label{Appendix_A_E11}
\omega^2\Bigl(1-3\frac{v_\mathrm{T}^2}{V^2}\Bigr)=\omega_\mathrm{p}^2,\qquad
\omega^2\Bigl(1-3\frac{v_\mathrm{T}^2}{V_\mathrm{r}^2}\Bigr)=\omega_{\mathrm{p}0}^2\,.
\end{equation}

Define

\begin{equation}\label{Appendix_A_E12}
\xi^2=\frac{1}{1-3\frac{v_\mathrm{T}^2}{V_\mathrm{r}^2}},\qquad \omega^2=\xi^2\omega_{\mathrm{p}0}^2\,.
\end{equation}

The condition \(u<V\) is satisfied when \(n_{e} < n_{e}(x)\). Thus one can calculate the probability

\begin{equation}\label{Appendix_A_E13}
P(u<V)=\int_{0}^{n_{e}(x)}P(n)\,dn\,.
\end{equation}

For a Gaussian distribution one can obtain the distribution in terms of the incomplete Gamma function (or, equivalently, in terms of error functions):

\begin{equation}\label{Appendix_A_E16}
P\bigl(u_i<V\bigr)
=\frac{1}{2}\Biggl\{1+\operatorname{sign}\Bigl(V^2-V_\mathrm{r}^2\Bigr)
\,\Gamma\!\Biggl(\frac{1}{2},\Biggl[\frac{n_\mathrm{e}}{\delta n}\frac{3\Bigl(\frac{v_\mathrm{T}^2}{V_\mathrm{r}^2}-\frac{v_\mathrm{T}^2}{V^2}\Bigr)}
{1-3\frac{v_\mathrm{T}^2}{V_\mathrm{r}^2}}\Biggr]^2\Biggr)\Biggr\}\,.
\end{equation}

Similarly, the probability that the velocity on the other end exceeds \(V\) is

\begin{equation}\label{Appendix_A_E17}
P\bigl(u_{i+1}>V\bigr)
=\frac{1}{\delta n\sqrt{\pi}}
\int_{\,\frac{1-3\frac{v_\mathrm{T}^2}{V^2}}{1-3\frac{v_\mathrm{T}^2}{V_\mathrm{r}^2}}\,n_\mathrm{e}}^{\,\frac{n_\mathrm{e}}{1-3\frac{v_\mathrm{T}^2}{V_\mathrm{r}^2}}}P(n)dn\,.
\end{equation}

that may be written as:

\begin{equation}\label{Appendix_A_E18}
P\bigl(u_{i+1}>V\bigr)
=\frac{1}{2\sqrt{\pi}}\Biggl\{
\Gamma\!\Biggl[\frac{1}{2},\Biggl(\frac{n_\mathrm{e}}{\delta n}\frac{3\frac{v_\mathrm{T}^2}{V_\mathrm{r}^2}}{1-3\frac{v_\mathrm{T}^2}{V_\mathrm{r}^2}}\Biggr)^2\Biggr]
-\operatorname{sign}\Bigl(V^2-V_\mathrm{r}^2\Bigr)
\Gamma\!\Biggl[\frac{1}{2},\Biggl(\frac{3\frac{v_\mathrm{T}^2}{V_\mathrm{r}^2}-3\frac{v_\mathrm{T}^2}{V^2}}{1-3\frac{v_\mathrm{T}^2}{V_\mathrm{r}^2}}\frac{n_\mathrm{e}}{\delta n}\Biggr)^2\Biggr]
\Biggr\}\,.
\end{equation}

The probability of reflection corresponds to the condition

\begin{equation}\label{Appendix_A_E19}
n > n_\mathrm{e}\,\xi^2 = \frac{n_\mathrm{e}}{1-3\frac{v_\mathrm{T}^2}{V_\mathrm{r}^2}}\,.
\end{equation}

Thus,

\begin{equation}\label{Appendix_A_E21}
P_\mathrm{ref}(\omega)=\frac{1}{\sqrt{\pi}}
\int_{\frac{n_\mathrm{e}}{1-3\frac{v_\mathrm{T}^2}{V_\mathrm{r}^2}}-n_\mathrm{e}}^{\infty}
\exp\!\left[-\Bigl(\frac{\delta n}{\delta n}\Bigr)^2\right]d\Bigl(\frac{\delta n}{\delta n}\Bigr)
=\frac{1}{2\sqrt{\pi}}\Biggl[
\Gamma\Bigl(\frac{1}{2}\Bigr)-\Gamma\Biggl(\frac{1}{2},\Bigl(\frac{n_\mathrm{e}}{\delta n}\frac{3\frac{v_\mathrm{T}^2}{V_\mathrm{r}^2}}{1-3\frac{v_\mathrm{T}^2}{V_\mathrm{r}^2}}\Bigr)^2\Biggr)
\Biggr]\,.
\end{equation}

In summary, the probability distribution for \(V\) for is given by
the case of multiple (infinite) number of reflections reads as
\begin{equation}\label{Appendix_A_E22}
P_{\omega}(V)=\frac{P\bigl(u_{i+1}>V\bigr)\cdot P\bigl(u_i<V\bigr)}
{1+P_\mathrm{ref}(\omega)}\,.
\end{equation}

i.e.,

\begin{align}\label{Appendix_A_E23}
P_{\omega}(V)
&= \frac{1}{2\pi}\,
\frac{%
  \Biggl\{
    \Gamma\!\Bigl[\tfrac{1}{2},
      \Bigl(\tfrac{n_\mathrm{e}}{\delta n}\,\tfrac{3v_\mathrm{T}^2}{V_\mathrm{r}^2}\Bigr)^2
    \Bigr]
    -\,
    \operatorname{sign}\!\Bigl(V^2 - V_\mathrm{r}^2\Bigr)\,
    \Gamma\!\Bigl[\tfrac{1}{2},
      \Bigl(\Bigl(\tfrac{v_\mathrm{T}^2}{V_\mathrm{r}^2}
               - \tfrac{v_\mathrm{T}^2}{V^2}\Bigr)\,
            \tfrac{3n_\mathrm{e}}{\delta n}
      \Bigr)^2
    \Bigr]
  \Biggr\}
}{
  1 \;+\; \tfrac{1}{2\sqrt{\pi}}
  \Biggl[
    \Gamma\!\Bigl(\tfrac{1}{2}\Bigr)
    -\,
    \Gamma\!\Bigl(\tfrac{1}{2},
      \Bigl(\tfrac{n_\mathrm{e}}{\delta n}\,\tfrac{3v_\mathrm{T}^2}{V_\mathrm{r}^2}\Bigr)^2
    \Bigr)
  \Biggr]
}
\nonumber\\[2mm]
&\quad \times
\Biggl\{
    \Gamma\!\Bigl(\tfrac{1}{2}\Bigr)
    +\,
    \operatorname{sign}\!\Bigl(V^2 - V_\mathrm{r}^2\Bigr)\,
    \Gamma\!\Bigl[\tfrac{1}{2},
      \Bigl(\tfrac{3n_\mathrm{e}}{\delta n}\,
            \Bigl(\tfrac{v_\mathrm{T}^2}{V_\mathrm{r}^2}
                 - \tfrac{v_\mathrm{T}^2}{V^2}\Bigr)
      \Bigr)^2
    \Bigr]
\Biggr\}\,.
\end{align}

An equivalent formulation in terms of error functions has  the following form:

\begin{equation}\label{Appendix_A_E24}
P\bigl(u_i<V\bigr)=\frac{1}{2}\Biggl\{1+\operatorname{erf}\!\Biggl[\frac{n_\mathrm{e}}{\delta n}\frac{3\Bigl(\frac{v_\mathrm{T}^2}{V_\mathrm{r}^2}-\frac{v_\mathrm{T}^2}{V^2}\Bigr)}{1-3\frac{v_\mathrm{T}^2}{V_\mathrm{r}^2}}\Biggr]\Biggr\}\,.
\end{equation}

and similarly,

\begin{equation}\label{Appendix_A_E25}
P\bigl(u_{i+1}>V\bigr)
=\frac{1}{2}\Biggl\{
\operatorname{erf}\!\Biggl[\frac{n_\mathrm{e}}{\delta n}\frac{3\frac{v_\mathrm{T}^2}{V_\mathrm{r}^2}}{1-3\frac{v_\mathrm{T}^2}{V_\mathrm{r}^2}}\Biggr]
-\operatorname{erf}\!\Biggl[\frac{n_\mathrm{e}}{\delta n}\frac{3\Bigl(\frac{v_\mathrm{T}^2}{V_\mathrm{r}^2}-\frac{v_\mathrm{T}^2}{V^2}\Bigr)}{1-3\frac{v_\mathrm{T}^2}{V_\mathrm{r}^2}}\Biggr]
\Biggr\}\,.
\end{equation}

and for the reflection probability,

\begin{equation}\label{Appendix_A_E26}
P_\mathrm{ref}(\omega)
=\frac{1}{2}\Biggl[1-\operatorname{erf}\!\Biggl(\frac{n_\mathrm{e}}{\delta n}\frac{3\frac{v_\mathrm{T}^2}{V_\mathrm{r}^2}}{1-3\frac{v_\mathrm{T}^2}{V_\mathrm{r}^2}}\Biggr)\Biggr]\,.
\end{equation}

Thus, the overall probability distribution may also be written as

\begin{equation}\label{Appendix_A_E27}
P_{\omega}(V)=\frac{1}{4}\,\frac{
\Biggl\{
\operatorname{erf}\!\Biggl(\frac{n_\mathrm{e}}{\delta n}\frac{3v_\mathrm{T}^2}{V_\mathrm{r}^2}\Biggr)
-\operatorname{erf}\!\Biggl[\frac{n_\mathrm{e}}{\delta n}\,3\Bigl(\frac{v_\mathrm{T}^2}{V_\mathrm{r}^2}-\frac{v_\mathrm{T}^2}{V^2}\Bigr)\Biggr]
\Biggr\}
\,
\Biggl\{
1+\operatorname{erf}\!\Biggl[3\frac{n_\mathrm{e}}{\delta n}\Bigl(\frac{v_\mathrm{T}^2}{V_\mathrm{r}^2}-\frac{v_\mathrm{T}^2}{V^2}\Bigr)\Biggr]
\Biggr\}
}{
1+\frac{1}{2}\Biggl[1-\operatorname{erf}\!\Biggl(\frac{n_\mathrm{e}}{\delta n}\frac{3v_\mathrm{T}^2}{V_\mathrm{r}^2}\Biggr)\Biggr]
}\,.
\end{equation}

\section{Growth and damping calculation} \label{Appendix_C}

\bigskip 

\begin{align}
    F_{b}(v,L,t) =
\begin{cases}
n_{b}(v) \dfrac{(\alpha -1)}{v_{\min}} \left( \dfrac{v_{\min}}{v} \right)^{\alpha}, & \text{for } v \geq \dfrac{L}{t}, \\
0, & \text{for } v < \dfrac{L}{t}.
\end{cases}
\end{align}

The expression for increment reads:

\begin{equation}
    \gamma = \pi \omega_\mathrm{p} \dfrac{n_\mathrm{b}}{n_\mathrm{e}} \int\limits_{0}^{\infty} V^{2} \dfrac{\partial F}{\partial V} P_{\omega}(V) \, dV
\end{equation}

Derivative  $\dfrac{\partial F}{\partial v}$ (without the multiplier $\pi \omega_\mathrm{p} \dfrac{n_\mathrm{b}}{n_\mathrm{e}}$) reads:

\begin{align}
    \dfrac{\partial F}{\partial v} =
\begin{cases}
\dfrac{\partial}{\partial v} \left[ \dfrac{(\alpha -1)}{v_{\min}} \left( \dfrac{v_{\min}}{v} \right)^{\alpha} \right], & \text{for } v \geq \dfrac{L}{t}, \\
0, & \text{for } v < \dfrac{L}{t}.
\end{cases}
\end{align}

Thus the increment:

\begin{align}
\gamma &=   \nu +\Gamma\nonumber \\
&= \pi \omega_\mathrm{p} \dfrac{n_\mathrm{b}}{n_\mathrm{e}} \times \left\{ -\alpha (\alpha -1) \int\limits_{U}^{\infty} \left( \dfrac{v_{\min}}{V} \right)^{\alpha -1} P_{\omega}(V) \, dV \right. \nonumber \\
&\quad \left. + U^{2} P_{\omega}(U) \dfrac{(\alpha -1)}{v_{\min}} \left( \dfrac{v_{\min}}{U} \right)^{\alpha} \right\}
\label{eq:gamma_increment}
\end{align}

\bigskip 

For the sake of simplicity let $P_{\omega}(V)$, be Gaussian:

\[
P_{\omega}(V) = \dfrac{1}{\sqrt{\pi} \Delta V} \exp \left[ -\dfrac{(V - V_\mathrm{r})^2}{(\Delta V)^2} \right]
\]

To determine the maximum let us calculate the derivative. Parameters of our study are $\alpha$, $v_{\min}$, and for Gaussian $P_{\omega}(V)$, parameters $V_\mathrm{r}$ and $\Delta V$. The variable is $U = \dfrac{L}{t}$.
After some simple but cumbersome calculations one can find 

\begin{equation}
\dfrac{\partial \gamma}{\partial U} = 2 (\alpha -1) \pi \omega_\mathrm{p} \dfrac{n_\mathrm{b}}{n_\mathrm{e}} \times \left[ 1 - \dfrac{U(U - V_\mathrm{r})}{\Delta V^{2}} \right] P_{\omega}(U) \left( \dfrac{v_{\min}}{U} \right)^{\alpha -1} 
\end{equation}

For maximum, the derivative should be equal to zero, thus:

\[
\left( \dfrac{v_{\min}}{U} \right) - \dfrac{(U - V_\mathrm{r}) v_{\min}}{\Delta V^{2}} = 0 
\]
The extremal points are determined by

\[
U_{1} = \dfrac{V_\mathrm{r}}{2} - \sqrt{ \dfrac{V_\mathrm{r}^{2}}{4} + \Delta V^{2} } \simeq -\dfrac{ \Delta V^{2} }{V_\mathrm{r}} 
\]

\[
U_{2} = \dfrac{V_\mathrm{r}}{2} + \sqrt{ \dfrac{V_\mathrm{r}^{2}}{4} + \Delta V^{2} } 
\]

As $\Delta V^{2}$ is supposed to be much less than $V_\mathrm{r}^{2}$, it may be written as:

\[
U_{2} = \dfrac{V_\mathrm{r}}{2} + \dfrac{V_\mathrm{r}}{2} \sqrt{1 + \dfrac{4 \Delta V^{2}}{V_\mathrm{r}^{2}}} \simeq V_\mathrm{r} \left( 1 + \dfrac{\Delta V^{2}}{V_\mathrm{r}^{2}} \right)
\]

Since $U$ is supposed to be positive, only $U_{2}$ represents a valid solution. This value should correspond to the maximum of the increment. Calculation of the second derivative confirms it. 
We suppose that the probability distribution has a sharp narrow maximum at $U = V_\mathrm{r}$.
This allows one to evaluate the increment at $U = V_\mathrm{r}$:

\[
\gamma =  \nu+\Gamma =
\]

\begin{align*}
&=\pi \omega_\mathrm{p} \dfrac{n_\mathrm{b}}{n_\mathrm{e}} \times \left\{ -\alpha (\alpha -1) \int\limits_{U}^{\infty} \left( \dfrac{v_{\min}}{V} \right)^{\alpha -1} P_{\omega}(V) \, dV \right. \\
&\quad \left. + U^{2} P_{\omega}(U) \dfrac{(\alpha -1)}{v_{\min}} \left( \dfrac{v_{\min}}{U} \right)^{\alpha} \right\}
\end{align*}

\bigskip

\[
\gamma(V_\mathrm{r}) = \Gamma + \nu = 
\]
\begin{align}
\pi \omega_\mathrm{p} \dfrac{n_\mathrm{b}}{n_\mathrm{e}} (\alpha -1) \times \left( \dfrac{v_{\min}}{V_\mathrm{r}} \right)^{\alpha -2} \nonumber\\
\times \left[ -\alpha \left( \dfrac{v_{\min}}{V_\mathrm{r}} \right) + \dfrac{1}{\sqrt{\pi}} \left( \dfrac{v_{\min}}{\Delta V} \right) \right]. \label{eq:final_equation}
\end{align}

It follows then the necessary condition for the instability to exist in the limit of Gaussian distribution for the phase velocity:

\begin{equation}
    \dfrac{1}{\sqrt{\pi}} \left( \dfrac{v_{\min}}{\Delta V} \right) > \alpha \left( \dfrac{v_{\min}}{V_\mathrm{r}} \right)
\end{equation}

which may be re-written as 

\begin{equation}
    V_\mathrm{r} > \alpha \sqrt{\pi} \Delta V.
\end{equation}

\software{None}

\bibliography{sample631}{}

\begin{thebibliography}{}
\expandafter\ifx\csname natexlab\endcsname\relax\def\natexlab#1{#1}\fi
\providecommand{\url}[1]{\href{#1}{#1}}
\providecommand{\dodoi}[1]{doi:~\href{http://doi.org/#1}{\nolinkurl{#1}}}
\providecommand{\doeprint}[1]{\href{http://ascl.net/#1}{\nolinkurl{http://ascl.net/#1}}}
\providecommand{\doarXiv}[1]{\href{https://arxiv.org/abs/#1}{\nolinkurl{https://arxiv.org/abs/#1}}}

\bibitem[{{Aubier} \& {Boischot}(1972)}]{Aubier72}
{Aubier}, M., \& {Boischot}, A. 1972, Astronomy and Astrophysics, 19, 343

\bibitem[{{Bale} {et~al.}(2016){Bale}, {Goetz}, {Harvey}, {Turin}, {Bonnell}, {Dudok de Wit}, {Ergun}, {MacDowall}, {Pulupa}, {Andre}, {Bolton}, {Bougeret}, {Bowen}, {Burgess}, {Cattell}, {Chandran}, {Chaston}, {Chen}, {Choi}, {Connerney}, {Cranmer}, {Diaz-Aguado}, {Donakowski}, {Drake}, {Farrell}, {Fergeau}, {Fermin}, {Fischer}, {Fox}, {Glaser}, {Goldstein}, {Gordon}, {Hanson}, {Harris}, {Hayes}, {Hinze}, {Hollweg}, {Horbury}, {Howard}, {Hoxie}, {Jannet}, {Karlsson}, {Kasper}, {Kellogg}, {Kien}, {Klimchuk}, {Krasnoselskikh}, {Krucker}, {Lynch}, {Maksimovic}, {Malaspina}, {Marker}, {Martin}, {Martinez-Oliveros}, {McCauley}, {McComas}, {McDonald}, {Meyer-Vernet}, {Moncuquet}, {Monson}, {Mozer}, {Murphy}, {Odom}, {Oliverson}, {Olson}, {Parker}, {Pankow}, {Phan}, {Quataert}, {Quinn}, {Ruplin}, {Salem}, {Seitz}, {Sheppard}, {Siy}, {Stevens}, {Summers}, {Szabo}, {Timofeeva}, {Vaivads}, {Velli}, {Yehle}, {Werthimer}, \& {Wygant}}]{BaleFIELDS}
{Bale}, S.~D., {Goetz}, K., {Harvey}, P.~R., {et~al.} 2016, \ssr, 204, 49, \dodoi{10.1007/s11214-016-0244-5}

\bibitem[{{Celnikier} {et~al.}(1983){Celnikier}, {Harvey}, {Jegou}, {Moricet}, \& {Kemp}}]{Celnikier83}
{Celnikier}, L.~M., {Harvey}, C.~C., {Jegou}, R., {Moricet}, P., \& {Kemp}, M. 1983, Astronomy and Astrophysics, 126, 293

\bibitem[{{Chen} {et~al.}(2013){Chen}, {Howes}, {Bonnell}, {Mozer}, {Klein}, \& {Bale}}]{ChenC13}
{Chen}, C.~H.~K., {Howes}, G.~G., {Bonnell}, J.~W., {et~al.} 2013, in American Institute of Physics Conference Series, Vol. 1539, Solar Wind 13, ed. G.~P. {Zank}, J.~{Borovsky}, R.~{Bruno}, J.~{Cirtain}, S.~{Cranmer}, H.~{Elliott}, J.~{Giacalone}, W.~{Gonzalez}, G.~{Li}, E.~{Marsch}, E.~{Moebius}, N.~{Pogorelov}, J.~{Spann}, \& O.~{Verkhoglyadova}, 143--146, \dodoi{10.1063/1.4811008}

\bibitem[{Chen {et~al.}(2024)Chen, Ma, Wu, Ning, Zhou, \& Bale}]{ChenL24}
Chen, L., Ma, B., Wu, D., {et~al.} 2024, The Astrophysical Journal Letters, 975, L37

\bibitem[{{Chen} {et~al.}(2021){Chen}, {Ma}, {Wu}, {Zhao}, {Tang}, \& {Bale}}]{ChenL21}
{Chen}, L., {Ma}, B., {Wu}, D., {et~al.} 2021, The Astrophysical Journal Letters, 915, L22, \dodoi{10.3847/2041-8213/ac0b43}

\bibitem[{{Dresing} {et~al.}(2021){Dresing}, {Warmuth}, {Effenberger}, {Klein}, {Musset}, {Glesener}, \& {Br{\"u}dern}}]{Dresing21}
{Dresing}, N., {Warmuth}, A., {Effenberger}, F., {et~al.} 2021, Astronomy and Astrophysics, 654, A92, \dodoi{10.1051/0004-6361/202141365}

\bibitem[{{Fox} {et~al.}(2016){Fox}, {Velli}, {Bale}, {Decker}, {Driesman}, {Howard}, {Kasper}, {Kinnison}, {Kusterer}, {Lario}, {Lockwood}, {McComas}, {Raouafi}, \& {Szabo}}]{Fox2016}
{Fox}, N.~J., {Velli}, M.~C., {Bale}, S.~D., {et~al.} 2016, \ssr, 204, 7, \dodoi{10.1007/s11214-015-0211-6}

\bibitem[{{Gerekos} {et~al.}(2024){Gerekos}, {Steinbr{\"u}gge}, {Jebaraj}, {Casillas}, {Donini}, {S{\'a}nchez-Cano}, {Lester}, {Magdaleni{\'c}}, {Peters}, {Romero-Wolf}, \& {Blankenship}}]{Gerekos24}
{Gerekos}, C., {Steinbr{\"u}gge}, G., {Jebaraj}, I.~C., {et~al.} 2024, Astronomy and Astrophysics, 683, A56, \dodoi{10.1051/0004-6361/202347900}

\bibitem[{{Ginzburg} \& {Zhelezniakov}(1958)}]{Ginzburg58}
{Ginzburg}, V.~L., \& {Zhelezniakov}, V.~V. 1958, \sovast, 2, 653

\bibitem[{{Hinkel-Lipsker} {et~al.}(1992){Hinkel-Lipsker}, {Fried}, \& {Morales}}]{Hinkel92}
{Hinkel-Lipsker}, D.~E., {Fried}, B.~D., \& {Morales}, G.~J. 1992, Physics of Fluids B, 4, 559, \dodoi{10.1063/1.860255}

\bibitem[{{Jebaraj} {et~al.}(2023{\natexlab{a}}){Jebaraj}, {Krasnoselskikh}, {Pulupa}, {Magdalenic}, \& {Bale}}]{Jebaraj23c}
{Jebaraj}, I.~C., {Krasnoselskikh}, V., {Pulupa}, M., {Magdalenic}, J., \& {Bale}, S.~D. 2023{\natexlab{a}}, The Astrophysical Journal Letters, 955, L20, \dodoi{10.3847/2041-8213/acf857}

\bibitem[{{Jebaraj} {et~al.}(2024){Jebaraj}, {Krasnoselskikh}, {Voshchepynets}, {Dudok de Wit}, {Larosa}, {Chen}, {Balikhin}, \& {Bale}}]{Jebaraj_AGU2024}
{Jebaraj}, I.~C., {Krasnoselskikh}, V., {Voshchepynets}, A., {et~al.} 2024, in AGU Fall Meeting Abstracts, Vol. 2024, SH21E--2882

\bibitem[{{Jebaraj} {et~al.}(2023{\natexlab{b}}){Jebaraj}, {Magdalenic}, {Krasnoselskikh}, {Krupar}, \& {Poedts}}]{Jebaraj23}
{Jebaraj}, I.~C., {Magdalenic}, J., {Krasnoselskikh}, V., {Krupar}, V., \& {Poedts}, S. 2023{\natexlab{b}}, Astronomy and Astrophysics, 670, \dodoi{10.1051/0004-6361/202243494}

\bibitem[{{Jebaraj} {et~al.}(2023{\natexlab{c}}){Jebaraj}, {Kouloumvakos}, {Dresing}, {Warmuth}, {Wijsen}, {Palmroos}, {Gieseler}, {Marmyleva}, {Vainio}, {Krupar}, {Wiegelmann}, {Magdalenic}, {Schuller}, {Battaglia}, \& {Fedeli}}]{Jebaraj23b}
{Jebaraj}, I.~C., {Kouloumvakos}, A., {Dresing}, N., {et~al.} 2023{\natexlab{c}}, Astronomy and Astrophysics, 675, A27, \dodoi{10.1051/0004-6361/202245716}

\bibitem[{{Kaiser}(2005)}]{Kaiser05}
{Kaiser}, M.~L. 2005, Advances in Space Research, 36, 1483, \dodoi{10.1016/j.asr.2004.12.066}

\bibitem[{{Kellogg} \& {Horbury}(2005)}]{Kellog05}
{Kellogg}, P.~J., \& {Horbury}, T.~S. 2005, Annales Geophysicae, 23, 3765, \dodoi{10.5194/angeo-23-3765-2005}

\bibitem[{{Kouloumvakos} {et~al.}(2015){Kouloumvakos}, {Nindos}, {Valtonen}, {Alissandrakis}, {Malandraki}, {Tsitsipis}, {Kontogeorgos}, {Moussas}, \& {Hillaris}}]{Kouloumvakos15}
{Kouloumvakos}, A., {Nindos}, A., {Valtonen}, E., {et~al.} 2015, Astronomy and Astrophysics, 580, A80, \dodoi{10.1051/0004-6361/201424397}

\bibitem[{{Krafft} \& {Savoini}(2022)}]{Krafft22a}
{Krafft}, C., \& {Savoini}, P. 2022, The Astrophysical Journal Letters, 924, L24, \dodoi{10.3847/2041-8213/ac46a7}

\bibitem[{{Krafft} \& {Volokitin}(2016)}]{Krafft16}
{Krafft}, C., \& {Volokitin}, A.~S. 2016, The Astrophysical Journal, 821, 99, \dodoi{10.3847/0004-637X/821/2/99}

\bibitem[{{Krafft} {et~al.}(2013){Krafft}, {Volokitin}, \& {Krasnoselskikh}}]{Krafft13}
{Krafft}, C., {Volokitin}, A.~S., \& {Krasnoselskikh}, V.~V. 2013, The Astrophysical Journal, 778, 111, \dodoi{10.1088/0004-637X/778/2/111}

\bibitem[{{Krafft} {et~al.}(2015){Krafft}, {Volokitin}, \& {Krasnoselskikh}}]{Krafft15}
---. 2015, The Astrophysical Journal, 809, 176, \dodoi{10.1088/0004-637X/809/2/176}

\bibitem[{{Krasnoselskikh} {et~al.}(2019){Krasnoselskikh}, {Voshchepynets}, \& {Maksimovic}}]{Krasnoselskikh19}
{Krasnoselskikh}, V., {Voshchepynets}, A., \& {Maksimovic}, M. 2019, The Astrophysical Journal, 879, 51, \dodoi{10.3847/1538-4357/ab22bf}

\bibitem[{{Krucker} {et~al.}(2007){Krucker}, {Kontar}, {Christe}, \& {Lin}}]{Krucker07}
{Krucker}, S., {Kontar}, E.~P., {Christe}, S., \& {Lin}, R.~P. 2007, The Astrophysical Journal Letters, 663, L109, \dodoi{10.1086/519373}

\bibitem[{{Lemaire}(2010)}]{Lemaire2010}
{Lemaire}, J. 2010, in American Institute of Physics Conference Series, Vol. 1216, Twelfth International Solar Wind Conference, ed. M.~{Maksimovic}, K.~{Issautier}, N.~{Meyer-Vernet}, M.~{Moncuquet}, \& F.~{Pantellini} (AIP), 8--13, \dodoi{10.1063/1.3395971}

\bibitem[{{Lin} {et~al.}(1986){Lin}, {Levedahl}, {Lotko}, {Gurnett}, \& {Scarf}}]{Lin86}
{Lin}, R.~P., {Levedahl}, W.~K., {Lotko}, W., {Gurnett}, D.~A., \& {Scarf}, F.~L. 1986, The Astrophysical Journal, 308, 954, \dodoi{10.1086/164563}

\bibitem[{{Lin} {et~al.}(1981){Lin}, {Potter}, {Gurnett}, \& {Scarf}}]{Lin81}
{Lin}, R.~P., {Potter}, D.~W., {Gurnett}, D.~A., \& {Scarf}, F.~L. 1981, The Astrophysical Journal, 251, 364, \dodoi{10.1086/159471}

\bibitem[{{Lorfing} {et~al.}(2023){Lorfing}, {Reid}, {G{\'o}mez-Herrero}, {Maksimovic}, {Nicolaou}, {Owen}, {Rodriguez-Pacheco}, {Ryan}, {Trotta}, \& {Verscharen}}]{Lorfing23}
{Lorfing}, C.~Y., {Reid}, H. A.~S., {G{\'o}mez-Herrero}, R., {et~al.} 2023, The Astrophysical Journal, 959, 128, \dodoi{10.3847/1538-4357/ad0be3}

\bibitem[{{Maksimovic} {et~al.}(1997){Maksimovic}, {Pierrard}, \& {Lemaire}}]{Lemaire97}
{Maksimovic}, M., {Pierrard}, V., \& {Lemaire}, J.~F. 1997, \aap, 324, 725

\bibitem[{{Melrose}(1970)}]{Melrose70}
{Melrose}, D.~B. 1970, Australian Journal of Physics, 23, 871

\bibitem[{{Melrose}(2017)}]{Melrose17}
---. 2017, Reviews of Modern Plasma Physics, 1, 5, \dodoi{10.1007/s41614-017-0007-0}

\bibitem[{{Mitchell} {et~al.}(2025){Mitchell}, {Christian}, {de Nolfo}, {Cohen}, {Hill}, {Kouloumvakos}, {Labrador}, {Leske}, {McComas}, {McNutt}, {Mitchell}, {Shen}, {Schwadron}, {Wiedenbeck}, {Bale}, \& {Pulupa}}]{Mitchell25}
{Mitchell}, J.~G., {Christian}, E.~R., {de Nolfo}, G.~A., {et~al.} 2025, The Astrophysical Journal, 980, 96, \dodoi{10.3847/1538-4357/adaa7c}

\bibitem[{{Mozer} {et~al.}(2024){Mozer}, {Agapitov}, {Bale}, {Goetz}, {Krasnoselskikh}, {Pulupa}, {Sauer}, \& {Voshchepynets}}]{Mozer24}
{Mozer}, F.~S., {Agapitov}, O., {Bale}, S.~D., {et~al.} 2024, Astronomy and Astrophysics, 690, L6, \dodoi{10.1051/0004-6361/202451134}

\bibitem[{Ogilvie {et~al.}(1977)Ogilvie, von Rosenvinge, \& Durney}]{Ogilve77}
Ogilvie, K.~W., von Rosenvinge, T., \& Durney, A.~C. 1977, Science, 198, 131, \dodoi{10.1126/science.198.4313.131}

\bibitem[{{Pierrard} \& {Lemaire}(1996)}]{Lemaire96}
{Pierrard}, V., \& {Lemaire}, J. 1996, \jgr, 101, 7923, \dodoi{10.1029/95JA03802}

\bibitem[{{Pulupa} {et~al.}(2024){Pulupa}, {Bale}, {Jebaraj}, {Romeo}, \& {Krucker}}]{Pulupa24}
{Pulupa}, M., {Bale}, S.~D., {Jebaraj}, I.~C., {Romeo}, O., \& {Krucker}, S. 2024, arXiv e-prints, arXiv:2412.05464, \dodoi{10.48550/arXiv.2412.05464}

\bibitem[{{Pulupa} {et~al.}(2017){Pulupa}, {Bale}, {Bonnell}, {Bowen}, {Carruth}, {Goetz}, {Gordon}, {Harvey}, {Maksimovic}, {Mart{\'\i}nez-Oliveros}, {Moncuquet}, {Saint-Hilaire}, {Seitz}, \& {Sundkvist}}]{Pulupa17}
{Pulupa}, M., {Bale}, S.~D., {Bonnell}, J.~W., {et~al.} 2017, Journal of Geophysical Research (Space Physics), 122, 2836, \dodoi{10.1002/2016JA023345}

\bibitem[{{Pulupa} {et~al.}(2020){Pulupa}, {Bale}, {Badman}, {Bonnell}, {Case}, {de Wit}, {Goetz}, {Harvey}, {Hegedus}, {Kasper}, {Korreck}, {Krasnoselskikh}, {Larson}, {Lecacheux}, {Livi}, {MacDowall}, {Maksimovic}, {Malaspina}, {Mart{\'\i}nez Oliveros}, {Meyer-Vernet}, {Moncuquet}, {Stevens}, \& {Whittlesey}}]{Pulupa20}
{Pulupa}, M., {Bale}, S.~D., {Badman}, S.~T., {et~al.} 2020, The Astrophysical Journal Supplements, 246, 49, \dodoi{10.3847/1538-4365/ab5dc0}

\bibitem[{{Scudder}(1992)}]{Scudder92}
{Scudder}, J.~D. 1992, \apj, 398, 319, \dodoi{10.1086/171859}

\bibitem[{{Scudder} \& {Karimabadi}(2013)}]{Scudder13}
{Scudder}, J.~D., \& {Karimabadi}, H. 2013, \apj, 770, 26, \dodoi{10.1088/0004-637X/770/1/26}

\bibitem[{{Sishtla} {et~al.}(2023){Sishtla}, {Jebaraj}, {Pomoell}, {Magyar}, {Pulupa}, {Kilpua}, \& {Bale}}]{Sishtla23}
{Sishtla}, C.~P., {Jebaraj}, I.~C., {Pomoell}, J., {et~al.} 2023, The Astrophysical Journal Letters, 959, L33, \dodoi{10.3847/2041-8213/ad137e}

\bibitem[{{Steinberg} {et~al.}(1971){Steinberg}, {Aubier-Giraud}, {Leblanc}, \& {Boischot}}]{Steinberg71}
{Steinberg}, J.~L., {Aubier-Giraud}, M., {Leblanc}, Y., \& {Boischot}, A. 1971, Astronomy and Astrophysics, 10, 362

\bibitem[{{Strauss} {et~al.}(2020){Strauss}, {Dresing}, {Kollhoff}, \& {Br{\"u}dern}}]{Strauss20}
{Strauss}, R.~D., {Dresing}, N., {Kollhoff}, A., \& {Br{\"u}dern}, M. 2020, The Astrophysical Journal, 897, 24, \dodoi{10.3847/1538-4357/ab91b0}

\bibitem[{{Suzuki} \& {Dulk}(1985)}]{Suzuki85book}
{Suzuki}, S., \& {Dulk}, G.~A. 1985, in Solar Radiophysics: Studies of Emission from the Sun at Metre Wavelengths, ed. D.~J. {McLean} \& N.~R. {Labrum} (Cambridge University Press), 289--332

\bibitem[{{Tkachenko} {et~al.}(2021){Tkachenko}, {Krasnoselskikh}, \& {Voshchepynets}}]{Tkachenko21}
{Tkachenko}, A., {Krasnoselskikh}, V., \& {Voshchepynets}, A. 2021, The Astrophysical Journal, 908, 126, \dodoi{10.3847/1538-4357/abd2bd}

\bibitem[{{Volokitin} \& {Krafft}(2018)}]{Volokitin18}
{Volokitin}, A.~S., \& {Krafft}, C. 2018, The Astrophysical Journal, 868, 104, \dodoi{10.3847/1538-4357/aae7cc}

\bibitem[{{Voshchepynets} \& {Krasnoselskikh}(2015)}]{Voshchep15b}
{Voshchepynets}, A., \& {Krasnoselskikh}, V. 2015, Journal of Geophysical Research (Space Physics), 120, 10,139, \dodoi{10.1002/2015JA021705}

\bibitem[{{Voshchepynets} {et~al.}(2015){Voshchepynets}, {Krasnoselskikh}, {Artemyev}, \& {Volokitin}}]{Voshchep15a}
{Voshchepynets}, A., {Krasnoselskikh}, V., {Artemyev}, A., \& {Volokitin}, A. 2015, The Astrophysical Journal, 807, 38, \dodoi{10.1088/0004-637X/807/1/38}

\bibitem[{{Willes} {et~al.}(1996){Willes}, {Robinson}, \& {Melrose}}]{Willes1996}
{Willes}, A.~J., {Robinson}, P.~A., \& {Melrose}, D.~B. 1996, Physics of Plasmas, 3, 149, \dodoi{10.1063/1.871841}

\bibitem[{Zaitsev {et~al.}(1974)Zaitsev, Kunilov, Mityakov, \& Rapoport}]{Zaitsev74}
Zaitsev, V., Kunilov, M., Mityakov, N., \& Rapoport, V. 1974, Soviet Astronomy, 18, 147

\bibitem[{Zheleznyakov \& Zaitsev(1970)}]{Zheleznyakov70}
Zheleznyakov, V., \& Zaitsev, V. 1970, Soviet Astronomy, 14, 47

\end{thebibliography}
\bibliographystyle{aasjournal}

\end{document}